\documentclass{aa}
\usepackage{graphicx}
\usepackage{txfonts}
\begin{document}
\title{The effects of driving time scales on coronal heating in a stratified atmosphere}
\author{T. A. Howson \inst{1} \and I. De Moortel \inst{1,2}}
\institute{School of Mathematics and Statistics, University of St. Andrews, St. Andrews, Fife, KY16 9SS, U.K. \and Rosseland Centre for Solar Physics, University of Oslo, PO Box 1029  Blindern, NO-0315 Oslo, Norway}

\abstract{}
{We investigate the atmospheric response to coronal heating driven by random velocity fields with different characteristic time scales and amplitudes.}
{We conducted a series of three-dimensional magnetohydrodynamic simulations of random driving imposed on a gravitationally stratified model of the solar atmosphere. In order to understand differences between alternating current (AC) and direct current (DC) heating, we considered the effects of changing the characteristic time scales of the imposed velocities. We also investigated the effects of the magnitude of the velocity driving.}
{In all cases, complex foot point motions lead to a proliferation of current sheets and energy dissipation throughout the coronal volume. For a given driving amplitude, DC driving typically leads to a greater rate of energy injection when compared to AC driving. This ultimately leads to the formation of larger currents, increased heating rates and higher coronal temperatures in DC simulations. There is no difference in the spatial distribution of energy dissipation across simulations, however, energy release events in AC cases tend to be more frequent and last for less time than in DC cases. This results in more asymmetric temperature profiles for field lines heated by AC drivers. Higher velocity driving is associated with larger currents, higher temperatures and the corona occupying a larger fraction of the simulation volume. In all cases, the majority of heating is associated with small energy release events, which occur much more frequently than larger events.}  
{When combined with observational results that highlight a greater abundance of oscillatory power in lower frequency modes, these findings suggest that energy release in the corona is more likely to be driven by longer time scale motions. In the corona, AC and DC driving will occur concurrently and their effects remain difficult to isolate. The distribution of field line temperatures and the asymmetry of temperature profiles may reveal the frequency and longevity of energy release events and therefore the relative importance of AC and DC heating.}
\maketitle


\section{Introduction}\label{sec:introduction}
Despite extensive research over much of the past century, the specific processes which maintain the high temperatures of the solar corona remain poorly constrained. It is widely accepted that energy is injected into the Sun's atmosphere by the convective flows that are observed at the photosphere. However, the nature of the subsequent transmission and dissipation of this energy is still unclear. This is known as the coronal heating problem and research continues into the energy dissipation mechanism(s) and the associated mass cycle between different layers of the atmosphere \citep[see reviews by][]{Narain1990, Narain1996, Klimchuk2006, Parnell2012, Reale2014, Sakurai2017}.

The majority of the mechanisms proposed in response to the coronal heating problem fall into one of two broad categories, namely; AC heating and DC heating. This dichotomy arises according to the time scales of the velocity flows that drive energy into the corona. In particular, if the driving time scales are short (in relation to the Alfv\'en crossing time along a given coronal structure), then we will see AC heating \citep[e.g. see reviews by][]{Arregui2015, VanDoors2020} and if the driving time scales are long, then we will see DC heating \citep[e.g., see review by][]{WilmotSmith2015}. With regards to nomenclature, we note that AC and DC heating mechanisms are sometimes referred to as \emph{wave-based} and \emph{reconnection-based}, respectively. However, we note that MHD wave modes are able to drive magnetic reconnection \citep[e.g.][]{Sakai1984, McLaughlin2009, Howson2021} and magnetic reconnection can drive oscillations, shocks and wave heating \citep[e.g.][]{Barta2007, Hseih2009, Kigure2010}. For either AC or DC heating, due to the high magnetic and viscous Reynolds numbers $(\gg 1)$ that are expected for typical coronal conditions, energy must be transported to small scales in the magnetic and/or velocity fields before it can be efficiently dissipated as heat.

Several processes have been proposed to promote a transfer of energy to small scales. In the context of DC heating, these include braiding \citep[e.g.][]{Parker1972, VanBalle1988, Peter2004, WilmotSmith2011}, flux tube twisting \citep[][]{Bareford2013, Gordovskyy2016} and/or MHD avalanches \citep[e.g.][]{Hood2016, Reid2018}. Meanwhile, for AC heating, these processes include resonant absorption \citep[e.g.][]{Ionson1978, Davilla1987, Poedts1989, Ofman1998} and phase mixing \citep[e.g.][]{Hasegawa1974, Heyvaerts1983, Parker1991}. Moreover, many studies have highlighted the propensity of MHD waves to trigger the development of turbulent-like flows either through the interaction of counter-propagating modes \citep[e.g.][]{Iroshnikov1964, Matthaeus1999, VanBalle2011} or through the onset of dynamic instabilities \citep[e.g.][]{Browning1984, Terradas2008a, Antolin2014, Hillier2019}. For these cases, energy will cascade to smaller scales until it inevitably reaches the dissipation length scale whereupon it will be converted to heat. 

The dissipation mechanism notwithstanding, it remains unclear whether coronal oscillations carry sufficient energy to balance expected losses \citep[for example, see different estimates by.][]{Tomczyk2007, DePontieu2007, McIntosh2011, Thurgood2014, Srivastava2017}. In particular, in closed coronal loops, the interaction between an imposed driver with a reflected wave will result in the time-averaged energy injection rate for simple, sinusoidal wave drivers being too low unless the driver frequency is resonant, the amplitude is very large or transport coefficients are significantly enhanced above classical values \citep[e.g.][]{Howson2019, Howson2021, Prok2019}. Despite this, periodic motions with a wide range of frequencies have been detected \citep[e.g.][]{Morton2015, Morton2016}, and significantly for this study, there is much more power in low frequency modes (similar to DC models).

An important difference between many models of AC and DC heating arises according to whether the system will store energy in the background coronal magnetic field (rather than simply in the perturbed component). As MHD wave energy is typically equipartitioned between the kinetic and magnetic components (over the duration of the a wave period), the amount of energy in a typical AC heating model is constrained by the wave amplitudes, which can often be observed directly. However, DC heating models will typically induce an increase in the background magnetic energy which will ultimately allow a sustained Poynting flux of energy into the system. This is difficult to measure directly and thus leads to large uncertainties in the energy content of the corona. There is of course no reason why short time scale driving cannot lead to the storage of magnetic energy in the background field, and this study eliminates this difference to allow a fair comparison between AC and DC models. This approach was also followed in \citet{Howson2020} where similar driving was imposed on a potential coronal arcade. In this previous study, we found that DC driving produced greater energy injection rates, current formation and ultimately, higher temperatures.

Understanding the mass-cycle that results from the coupling of the corona to lower layers of the atmosphere remains a significant but important challenge. In particular, the small length scales that exist in the transition region present difficulties for accurately reproducing the mass flux into the corona during heating events \citep{Bradshaw2013}. Overcoming this problem with innovative numerical techniques \citep[e.g.][]{Lionello2009, Johnston2019, Johnston2021} is critical as, despite the low plasma-$\beta$ conditions in the corona, cooling times and any synthetic emission produced from simulations are highly sensitive to the evolution of the coronal density \citep[e.g.][]{Antiochos1978, Bradshaw2010, Bradshaw2011, Winebarger2018}. Furthermore, the efficiency of wave heating models critically depends on this density profile and it remains unclear whether the heating is able to self-consistently generate and/or maintain the density structuring \citep{Cargill2016, VanDamme2020}. 

In this paper we investigate the effects of driving time scales on energy release within a stratified atmosphere that includes the coupling between the corona and the chromosphere. In Sect. \ref{num_method}, we outline our model, including the initial conditions and the driving profiles. Then, in Sect. \ref{Sec_Res}, we present our results and, finally, in Sect. \ref{Discussion}, we discuss the implications of our findings in the context of coronal heating research.

\section{Numerical method} \label{num_method}
The numerical simulations presented within this article were conducted with the Lagrangian-Remap code, Lare3d \citep{Arber2001}. The scheme advances the full, resistive, three dimensional, MHD equations in normalised form. They are given by
\begin{equation}\frac{\text{D}\rho}{\text{D}t} = -\rho \vec{\nabla} \cdot \vec{v}, \end{equation}
\begin{equation} \label{eq:motion} \rho \frac{{\text{D}\vec{v}}}{{\text{D}t}} = \vec{j} \times \vec{B} - \vec{\nabla} P - \rho \vec{g} + \vec{F}_{\text{visc.}}, \end{equation}
\begin{equation} \label{eq:energy} \rho \frac{{\text{D}\epsilon}}{{\text{D}t}} = - P(\vec{\nabla} \cdot \vec{v}) - \vec{\nabla} \cdot \vec{F}_C - \rho^2\Lambda(T) + \eta \lvert \vec{j}\rvert^2 + Q_{\text{visc.}}, \end{equation}
\begin{equation}\label{eq:induction}\frac{\text{D}\vec{B}}{\text{D}t}=\left(\vec{B} \cdot \vec{\nabla}\right)\vec{v} - \left(\vec{\nabla} \cdot \vec{v} \right) \vec{B} - \vec{\nabla} \times \left(\eta \vec{\nabla} \times \vec{B}\right), \end{equation}
\begin{equation}\label{eq:state} P = 2 k_BnT.
\end{equation}
Here, $\rho$ is the plasma density, $\vec{v}$ is the velocity, $\vec{j}$ is the current density, $\vec{B}$ is the magnetic field, $\vec{g}$ is the acceleration due to gravity, $P$ is the gas pressure, $\epsilon$ is the specific internal energy density, $k_B$ is the Boltzmann constant and $n$ is the number density. In the energy equation (\ref{eq:energy}), the second and third terms on the right hand side correspond to thermal conduction and optically thin radiation, respsectively. In the conduction term, $\vec{F}_C$ is the Spitzer-H{\"a}rm heat flux. In the radiation term, $\Lambda(T)$, is a piecewise continuous function that approximates radiative losses in an optically thin plasma. This function is described in \citet{Klimchuk2008}. We impose a temperature floor of $2 \times 10^4$ K which prevents the plasma cooling below this temperature. In practice, this places a lower bound on the chromospheric temperature which is acceptable as we are not fully simulating the chromospheric physics (instead it acts as a mass reservoir).

In order to alleviate the difficulties of numerically resolving the conductive flux in the spatially under-resolved transition region, we employ a technique outlined in \citet{Linker2001, Lionello2009, Mikic2013}. In particular, adapted thermal conduction and radiative loss terms are imposed below a cutoff temperature, $T_c$, which we set to be $2.5 \times 10^4$ K. This has the effect of artificially broadening the transition region and has been shown to accurately simulate coronal temperature and density profiles (when compared to high resolution 1D models) in both static \citep{Lionello2009} and dynamic loops \citep{Mikic2013}.

In these equations, we have included the resistivity, $\eta$ and viscosity, $\nu$ as non-ideal terms which dissipate energy from the magnetic and velocity fields, respectively. The viscosity is a sum of contributions from two small shock viscosity terms which are included within all simulations to ensure numerical stability. Together, these contribute a force, $\vec{F}_{\text{visc.}}$ on the right-hand side of the equation of motion \ref{eq:motion} and a heating term, $Q_{\text{visc}}$ to the energy equation (\ref{eq:energy}). These terms are discussed in detail in \citet{Arber2018}. Using the notation detailed in the referenced manual, we have set $\text{visc1} = 0.05$ and $\text{visc2} =0.25$. 

As we are not considering the correct physical treatment of the chromosphere, we seek to only allow significant energy dissipation within the coronal portion of the domain (see below). As such, we set $\eta = 0$ in the volume close to the upper and lower boundaries. This has the added benefit of minimising the slippage of magnetic foot points through the imposed driving. More specifically, the resistivity is defined to be
\begin{equation}
\eta = \eta_0\left\{\tanh(z+\alpha) - \tanh(z-\alpha)\right\},
\label{eq:eta}
\end{equation}
where $\alpha = z_{\text{max}} - 9$ Mm and $\eta_0$ is a constant that yields a characteristic magnetic Reynolds number of $5 \times 10^3$ in the coronal volume (for the current numerical resolution and characterstic length scales). The scheme employed here does not enforce energy conservation and, thus, any temperature increase is not associated with numerical dissipation.

\begin{figure}[h]
  \centering
  \includegraphics[width=0.49\textwidth]{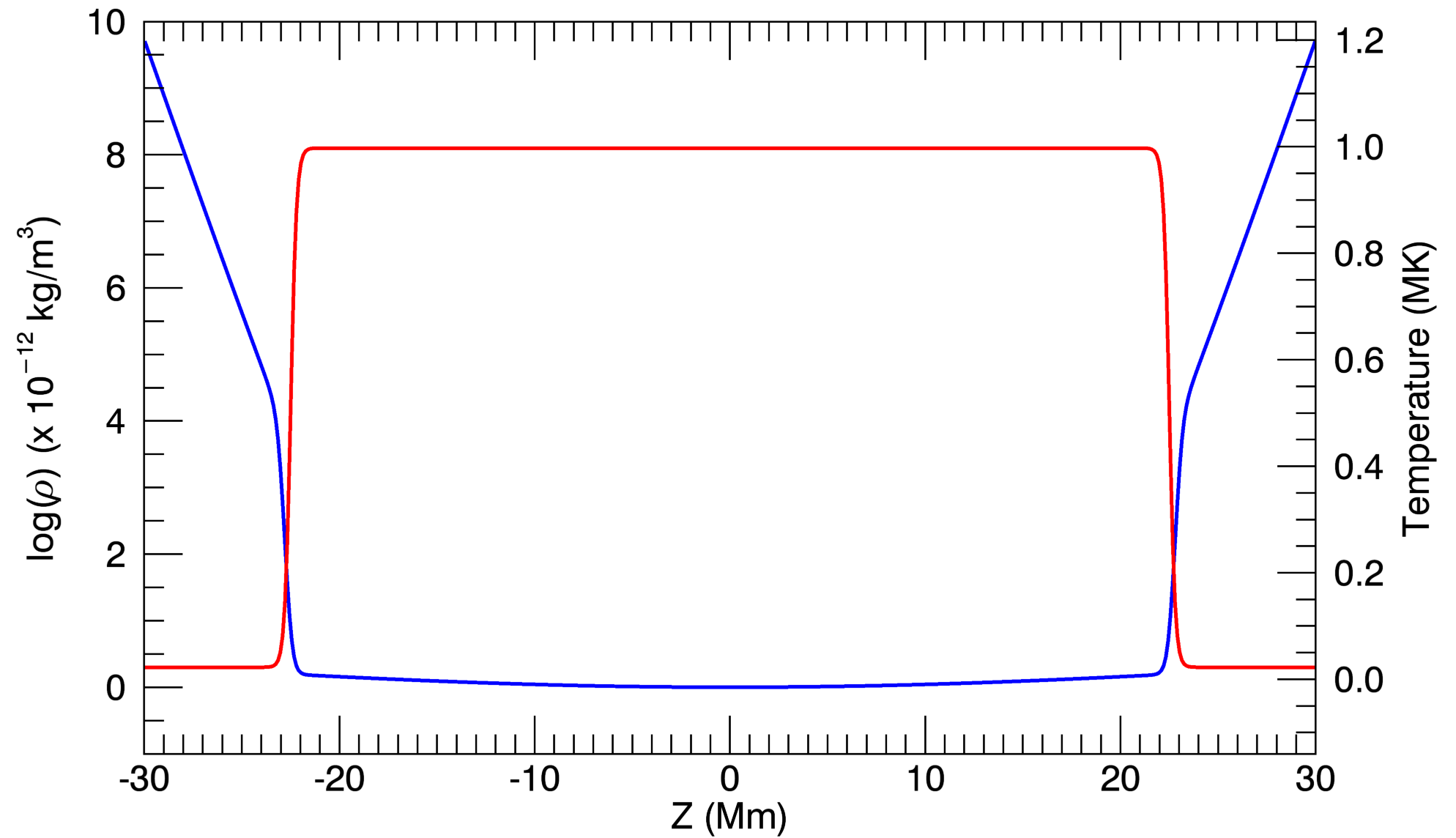}
  \caption{Initial density (blue) and temperature (red) profiles as a function of $z$. The simulations are initially invariant in the $x$ and $y$ directions.}
  \label{initial_cond}
\end{figure}

\subsection{Initial conditions}
For this study, we have simplified the geometry of the corona by modelling curved loops as straight magnetic structures which are embedded in a dense chromosphere at both foot points. The initial conditions consisted of a uniform magnetic field with a strength of 20 G. The field is initially parallel to the $z$-axis. We considered a numerical domain with dimensions of -10 Mm $\le x \le$ 10 Mm, -10 Mm $\le y \le$ 10 Mm and -30 Mm $\le z \le$ 30 Mm. The simulations used a numerical domain of $256 \times 256 \times 512$ grid points, in the $x, y$ and $z$ directions, respectively. In these simulations, we assume that the graviational acceleration, $\vec{g}$ (see equation \ref{eq:motion}) is field aligned, and is modified with height as would be expected in a semi-circular loop. In particular, the parallel component of gravity is given by
\begin{equation}
g_{\parallel} = g_0 \sin\left(\frac{\pi z}{2 z_{\text{max}}}\right) ,
\end{equation}
where $g_0 \approx 274 \text{ m s}^{-2}$ is the acceleration due to gravity at the solar surface. This acceleration is vertically downwards for $z<0$ and upwards (towards the loop foot point at z=$z_{\text{max}}$) for $z>0$. 

Our simulations begin with an imposed temperature profile, $T = T(z)$, given by
\begin{equation} \label{eq:temp}
T(z) = T_{\text{ch}}  + \frac{T_{\text{co}}  -  T_{\text{ch}}}{2} \left\{\tanh \left(\frac{z + a}{b}\right) - \tanh\left(\frac{z-a}{b}\right)\right\},
\end{equation}
where $T_{ch} = 2 \times 10^4$ K is the chromospheric temperature, $T_{\text{co}} = 10^6$ K is the initial coronal temperature, $a = 22.5$ Mm controls the location of the transition region and $b = 0.3$ Mm controls the width of the transition region. The temperature profile is displayed as the red line in Fig. \ref{initial_cond}.

We define initial conditions that are in hydrostatic equilibrium using 
\begin{equation}
\frac{\mathrm{d} P}{\mathrm{d}z} = - \rho g_{\parallel},
\end{equation}
to define the density, $\rho = \rho(z)$. Here we use equations (\ref{eq:state}) and (\ref{eq:temp}) to eliminate the pressure and temperature and then integrate numerically using the 4th order Runge-Kutta scheme. The logarithm of the resulting density profile is shown by the blue line in Fig. \ref{initial_cond}.

As there is no background heating term included in these conditions, the system is not in an initial thermodynamic equilibrium. In particular, the coronal volume will lose energy to the lower layers of the atmosphere through thermal conduction and all plasma above the temperature floor ($2 \times 10^4$ K) will cool due to optically thin radiative losses. Instead, we aim to see whether the driving can support a hot corona and how the long-term state differs for different imposed velocity profiles.

\begin{figure}[h]
  \centering
  \includegraphics[width=0.49\textwidth]{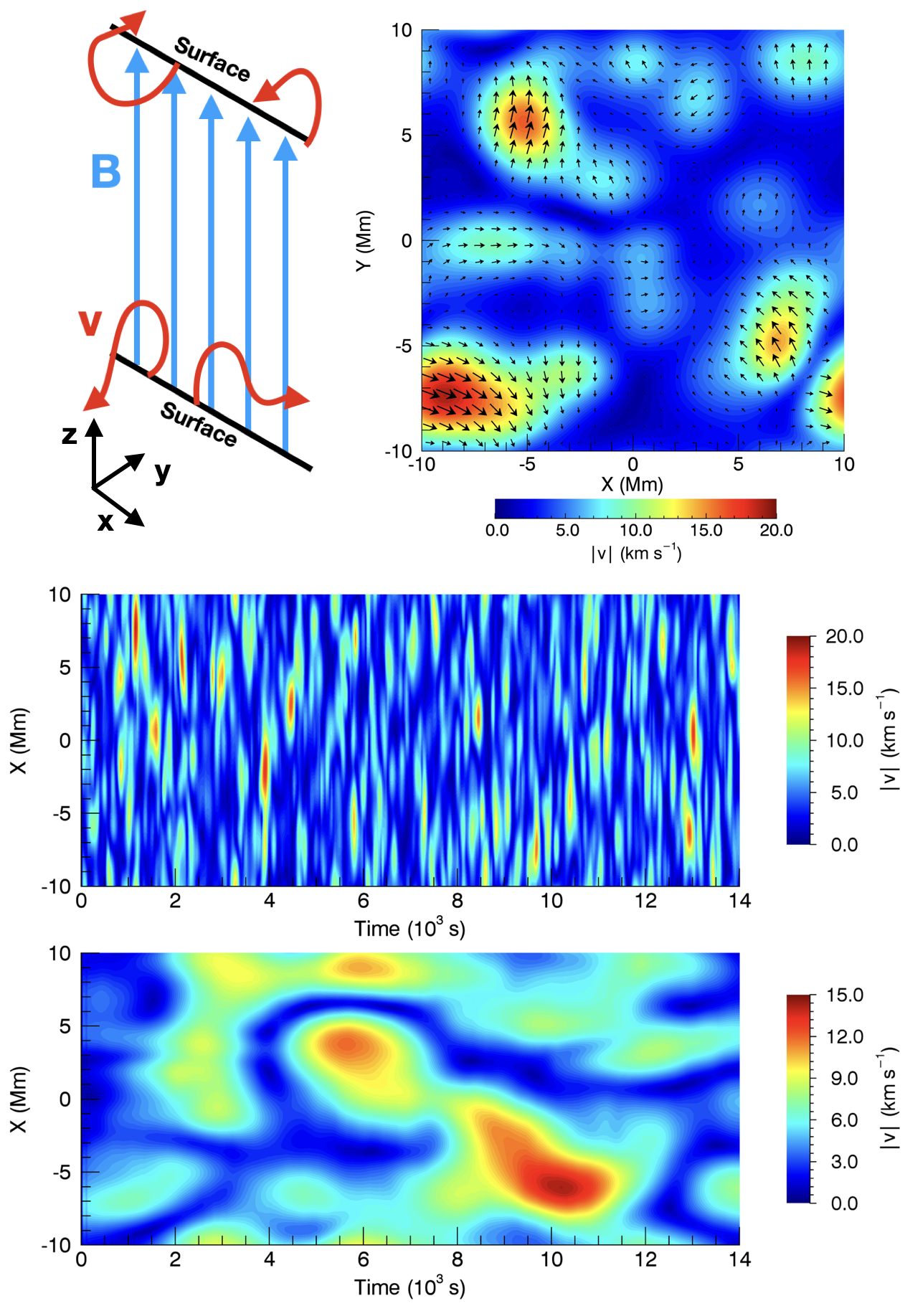}
  \caption{Imposed velocity drivers. \emph{Upper left:} Schematic of the drivers imposed on each of the foot points of magnetic field lines.  \emph{Upper right:} Imposed velocity field on the lower boundary of the medium amplitude AC simulation. \emph{Centre:} Temporal evolution of the magnitude of the imposed driver on the lower boundary of the medium amplitude AC simulation on the line $y=0$ Mm.  \emph{Lower:} Temporal evolution of the magnitude of the imposed driver on the lower boundary of the medium amplitude DC simulation on the line $y=0$ Mm.}
  \label{driver_plots}
\end{figure}

\subsection{Boundary conditions}
We model the convective flows that exist at the photosphere by imposing a transverse, space- and time-dependent velocity profile at both the upper and lower $z$ boundaries of the domain. We note that our simulations do not include the full-complexity of chromospheric physics and as such the exact manner in which energy is transmitted through the lower atmosphere is not considered here. As described in \citet{Howson2020}, we define a driver using a sum of many individual two-dimensional Gaussians, each of which has a particular amplitude, direction, length scale and time scale. In particular, on both photospheric boundaries ($z = \pm 30$ Mm), we impose $\vec{v} = (v_x, v_y, 0)$, where
\begin{eqnarray} \label{define_driver}
\label{vx_def} v_x = \displaystyle \sum_{i=1}^N v_i \cos{\theta_i} \exp\left\{\frac{-(r-r_i)^2}{l_i^2}\right\}\exp\left\{\frac{-(t-t_i)^2}{\tau_i^2}\right\},\\
\label{vy_def} v_y = \displaystyle \sum_{i=1}^N v_i \sin{\theta_i} \exp\left\{\frac{-(r-r_i)^2}{l_i^2}\right\}\exp\left\{\frac{-(t-t_i)^2}{\tau_i^2}\right\}.
\end{eqnarray}
Here, for each $i$ in the summation, $v_i$ is the amplitude of the velocity component, $\theta_i$ defines the direction of the component, $r_i$ is the centre of a two-dimensional Gaussian, $l_i$ is a parameter which defines the spatial scales of the velocity driver (the width of the 2D Gaussian), $t_i$ is the time of peak amplitude for each component and $\tau_i$ defines the time scale of each component. 

For each velocity component, all parameters are randomly selected from some statistical distribution. In particular, for all $i$, the $v_i$ are normally distributed with mean $v_{\mu}$ and variance $v^2_{\mu}/25$, the $\theta_i$ are uniformly distributed on the interval $\left[0, 2 \pi\right]$, the $r_i$ are uniformly distributed over the driven boundaries, the $l_i$ are normally distributed with mean $L/4 = 2.5$ Mm and variance $L^2/400 = 0.25 \text{ Mm}^2$, the $t_i$ are uniformly distributed over the duration of the simulation, and the $\tau_i$ are normally distributed with mean $\tau_{\mu}$ and variance $\tau^2_{\mu}/16$. We note that smaller values of $\tau_{\mu}$ create shorter time scales for the velocity driver (AC driving) and larger values are associated with longer time scales (DC driving). In equations \ref{vx_def} \& \ref{vy_def}, the number of terms included in the summation, $N$, is selected to be a function of the typical time scale $\tau_{\mu}$ and is chosen such that a similar number of components are active at all times. This ensures that the complexity of the velocity driver is consistent between different simulations in the parameter space. 

In this article, we conduct a parameter study on the mean amplitude, $v_{\mu}$ and the velocity time scale $\tau_{\mu}$. We consider three different amplitudes; low, medium and high, and two different characteristic time scales; AC and DC driving. The velocity drivers imposed in the low, medium and high amplitude simulations have a mean magnitude of approximately, 2, 4 and 6 $\text{ km s}^{-1}$, respectively. The velocity Gaussian components have an average duration of approximately 70 s in the AC simulations and approximately 1400 s in the DC simulations. For reference, the initial Alfv\'en crossing time between the two $z$ boundaries is approximately 450 s. 

In Fig. \ref{driver_plots}, we show the key features of the imposed driver. In the upper left hand panel, we show a schematic of the model with blue arrows representing the magnetic field and red arrows representing the action of the velocity drivers on the $z$ boundaries. In the upper right hand panel, we show a snapshot of the velocity field on the lower boundary of the medium amplitude AC simulation. In the lower two panels, we show the temporal evolution of the imposed velocity on the medium amplitude AC (centre) and DC (lower) simulations in the line $x=0$ Mm, $z = - 30$ Mm. We note that the velocity components have much longer lifetimes in the lower panel (DC) than in the central panel. Movies of AC and DC driving are included in the files that accompany this article.

In all simulations, the $x$ and $y$ boundaries are defined to be periodic. For the $z$ boundaries, with the exception of the imposed velocity drivers described above, zero-gradient conditions are enforced for all variables.

\section{Results}\label{Sec_Res}
We begin by describing the general evolution of the medium amplitude AC simulation. We initially focus on the general properties of the system that persist across all simulations. The specific differences that result from different velocity drivers will be detailed in subsequent sections. 

In all cases, the imposed driving at field line foot points stresses the magnetic field and induces the formation of currents throughout the domain. In the resistive subvolume (see equation \ref{eq:eta}), this then permits energy release through Ohmic heating and magnetic reconnection, leading to an increase in the plasma temperature. Conductive fronts then transfer this heat along field lines to the dense transition region and chromosphere. This generates an increase in the gas pressure and propels the evaporation of plasma into the coronal volume. The higher temperatures/densities enhance the conductive/radiative losses in the corona and, as a result, the temperature does not continuously increase. Although the driving, heating and cooling processes are inherently non-uniform in space and time, after a long time (in terms of the Alfv\'en travel time and the driving time scales), a statistically steady-state is obtained, where the coronal temperature, mass content and energy release rate remain relatively constant in time.

\begin{figure}[h]
  \centering
  \includegraphics[width=0.45\textwidth]{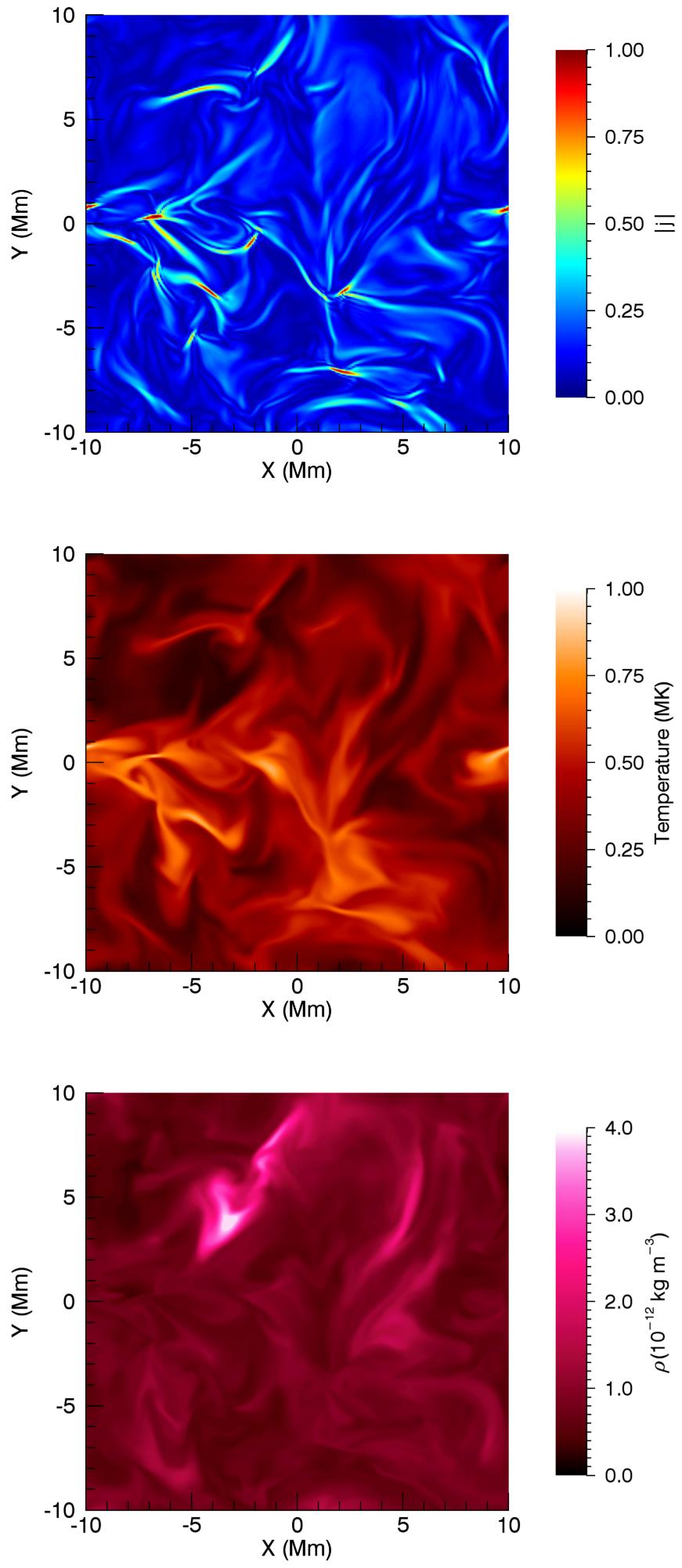}
  \caption{Horizontal cuts of the modulus of the current density, the temperature and the density at $t = 10900$ s in the medium amplitude AC simulation. We have normalised the current to the maximum value in this plane. The temporal evolution of this figure is shown in a movie that accompanies this article.}
  \label{horiz_contour}
\end{figure}

\begin{figure*}[h]
  \centering
  \includegraphics[width=\textwidth]{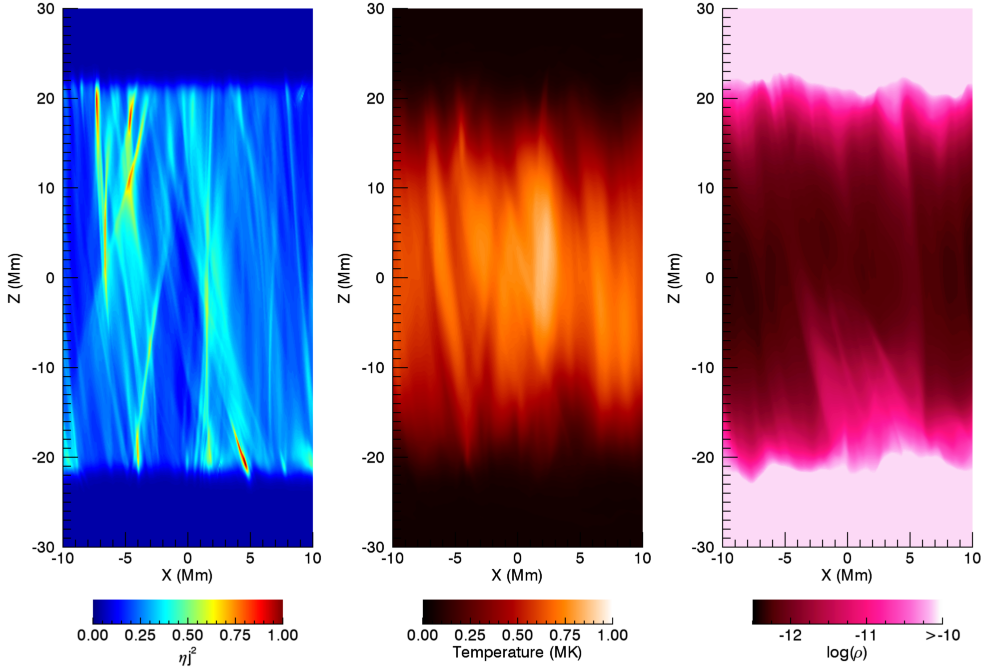}
  \caption{The Ohmic heating ($\eta j^2$; left), temperature (centre) and density (right) averaged in $y$ and shown as a function of $x$ and $z$. For the density, the logarithm is plotted and the colour bar is saturated at $10^{-10} \text{ kg m}^{-3}$ in order to show the details of coronal structures. The panels shown here correspond to the medium amplitude AC simulation.}
  \label{vert_contour}
\end{figure*}

In Fig. \ref{horiz_contour}, we show horizontal cuts of the modulus of the current density (upper), the temperature (centre) and the density (lower) at $z=0$ Mm (simulation midplane) at $t = 10900$ s (approx. 3 hours). The temporal evolution of these quantities is shown in the accompanying movie. We see that the currents form narrow sheets that are distributed throughout the entire cross-section. However, despite their spatial spread, the largest values are very localised, which leads to distinct heating events. In the movie, we see that these are also short-lived in time, producing a burtsy, impulsive heating profile even in this AC driving case.

In this resistive regime, the currents lead to energy release and an increase in the plasma temperature. As such, it is not surprising that the temperature profile (central panel of Fig. \ref{horiz_contour}) shows signatures of features that can be identified in the profile of $|\vec{j}|$. For example, we see a high current and temperature region in the area around $x = -5$ Mm, $y = -2$ Mm and similar ribbon structures at $x = -5$ Mm, $y=6$ Mm. However, it is clear that the temperature features tend to be more diffuse. There are several reasons for this. Firstly, as plasma does not cool instantaneously, the temperature will show regions of previously heated plasma, which are no longer associated with large currents in the upper panel. Secondly, thermal conduction plays an important role. Whilst only negligible amounts of energy are transferred across field lines, it will efficiently move heat along field lines. As such, the temperature profile can show signatures of heating (and currents) at different altitudes (that are not seen in the upper panel of Fig. \ref{horiz_contour}). Furthermore, at this time many field lines have a significant component parallel to this plane (due to the sustained driving), and conduction will therefore transfer heat across the plane. Finally, as the plasma-$\beta$ is non-zero, the heating will lead to cross-field pressure forces which will slightly modify the distribution of hot plasma.

The density profile (bottom panel of Fig. \ref{horiz_contour}) exhibits different structures to those observed in the other two panels. Indeed, there is some out-of-phase behaviour between the temperature and density panels. In particular, the high density region at $x = -3$ Mm, $y=4$ Mm, is at a low temperature. This is characteristic of the formation of a coronal condensation. As the plasma is at the field line apex, there is no gravitational force to pull the dense plasma towards either of the magnetic foot points. Therefore, the plasma can remain relatively stable at the midplane for a long time before eventually falling towards one of the foot points. The radiative losses here will be enhanced (due to the high density and cool temperature) and this increased cooling rate will further reduce the temperature, lowering the gas pressure and drawing more plasma into the condensation.

In Fig. \ref{vert_contour}, we display vertical profiles of the Ohmic heating rate ($\eta j^2$; left), the temperature (centre) and the density (right) for this simulation at the same time ($t = 10900$ s). For all three panels we have averaged along the $y$ axis to provide an insight into the nature of the variables throughout the whole domain. The mean Ohmic heating is normalised to the maximum value obtained during the simulation. For clarity, we plot the logarithm of the mean density and note that the chromospheric density is saturated to allow the coronal structure to be seen. The temporal evolution of this figure is shown in the accompanying movie.

For the Ohmic heating (left panel), the $\eta = 0$ regions are apparent for $|z| \gtrsim 21$ Mm. We note that large currents do develop within these regions but they do not contribute directly to plasma heating. We see that currents tend to form in long narrow structures that are aligned with magnetic field lines and that the highest Ohmic heating rates occur in small, intense bursts. These energy release events are also short-lived in time (see accompanying movie) and are reminiscent of nanoflares. 

In the central panel we see that a corona with a temperature of approximately 0.6 MK is maintained for this simulation (more energetic simulations are discussed in more detail below). The temperature of the chromosphere remains at $2 \times 10^4$ K, partly due to the lack of heating here but mainly due to the huge radiative losses that any heated plasma will experience here. As with the Ohmic heating profile, the temperature also shows narrow structuring that is aligned with the magnetic field. This is partly due to the nature of the heating but also due to thermal conduction spreading thermal energy along (and not across) magnetic field lines. We note that it is difficult to identify the signatures of individual heating events once the integration in the $y$ direction has been performed. 

In the right hand panel, we see that the heating in this simulation is sufficient to sustain a coronal density of approximately $10^{-12} \text{ kg m}^{-3}$ with some dense structures forming in response to the driving. The atmosphere remains well stratified and, once again, as plasma evaporation does not occur instantaneously with heating events, there is little connection between the density and the temperature/heating profiles. We note that the density enhancement of the visible structures in relation to the ambient corona often exceeds an order of magnitude. This is interesting in the context of wave heating models which typically require a significant density gradient across coronal flux tubes. However, in the accompanying movie, we note that the evolution of the density (as well as the other two variables) is highly dynamic. This is a consequence of the short time scaling driving considered in this simulation and it is unclear whether the density structuring is sufficiently long-lived to allow for classical wave heating mechanisms (e.g. phase mixing and resonant absorption) to release energy efficiently. Given these structures will have longer lifetimes in the DC simulations, a combination of AC and DC driving may produce classical wave heating processes (such as phase mixing and resonant absorption) in these structures. 

\subsection{Poynting flux}
The imposed drivers inject energy into the system at both the upper and lower $z$ boundaries of the domain. Since no perpendicular flows are permitted through the boundary, the only energy flux term is given by the Poynting flux. Meanwhile, there is an energy loss term that is associated with the radiative losses. Therefore, in terms of the volume integrated energy, $\Xi$, we have
\begin{equation}
\frac{\mathrm{d}\Xi}{\mathrm{d} t} + \frac{1}{\mu_0} \int \vec{E} \times \vec{B} \cdot \mathrm{d} \vec{S} = - \int R_L \,\, \mathrm{d} V.
\end{equation}
Here, the first integral represents the Poynting flux where, $\mu_0$ is the magnetic permeability in a vacuum, $\vec{E}$ is the electric field and the integral is calculated over both the upper and lower $z$ boundaries. The second term on the right-hand side represents the volume integrated radiative losses. Here we recall that the energy lost by plasma at $2 \times 10^4$ K is effectively zero (due to the inclusion of the temperature floor). For the current geometry, the energy injected through the driven boundaries can be expressed as \citep[e.g.][]{Parnell2012, Howson2020}
\begin{equation} \label{Poynt_flux_exp}
\int_{A_1} \left(v_xB_x + v_yB_y \right) B_z \,\, \mathrm{d} A \, - \int_{A_2}  \left(v_xB_x + v_yB_y \right) B_z \,\, \mathrm{d} A,
\end{equation}
where the two area integrals are calculated over the upper and lower boundaries respectively.

\begin{figure}[h]
  \centering
  \includegraphics[width=0.49\textwidth]{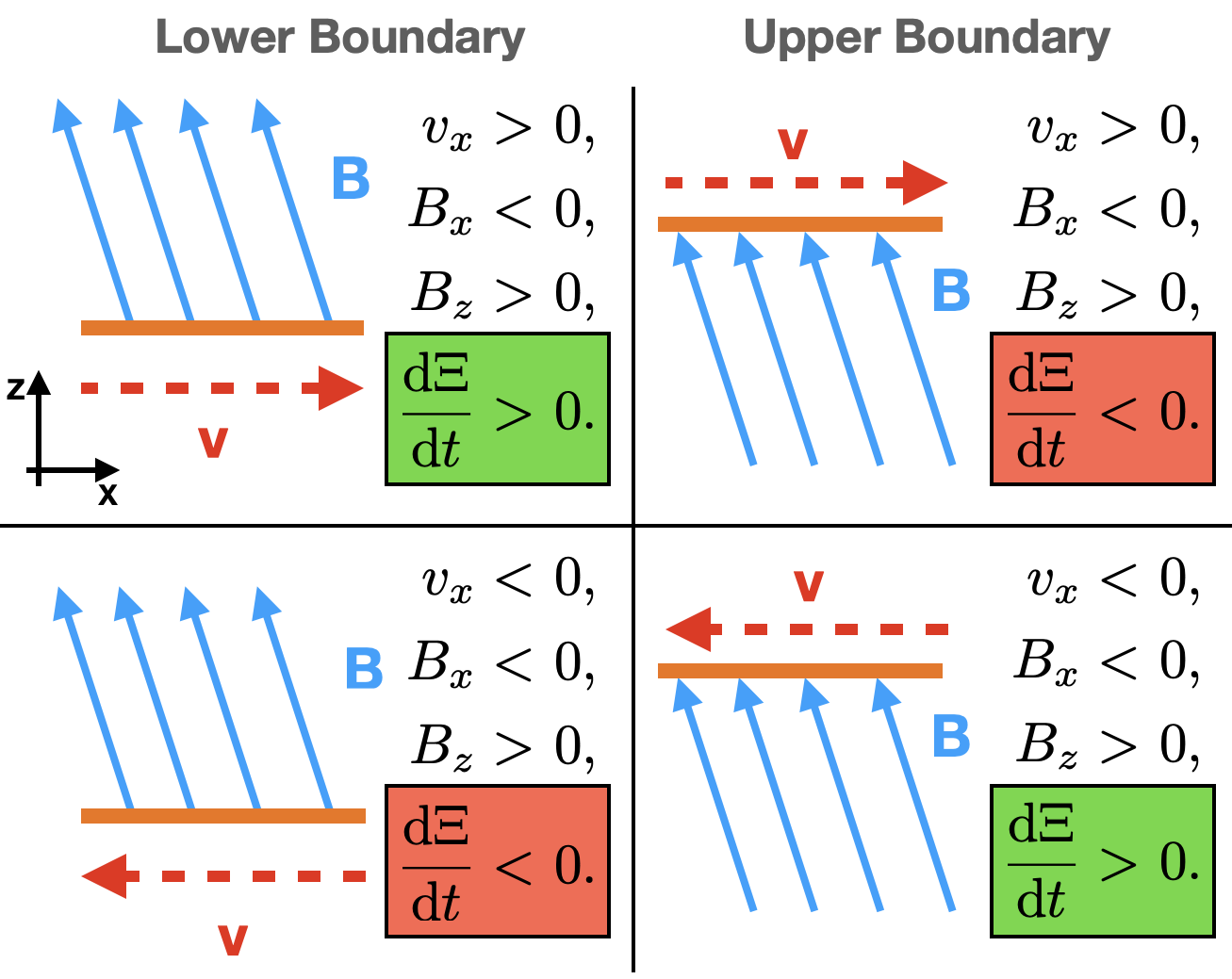}
  \caption{Schematic of the energetic consequences for different magnetic field and driving profiles. With the current geometry in mind, we show driving on the lower (left panels) and upper (right panels) $z$ boundaries.}
  \label{poynt_cartoon}
\end{figure}

In Fig. \ref{poynt_cartoon}, we show a schematic of the implications of equation (\ref{Poynt_flux_exp}) on different driving profiles. For simplicity, we consider a case with $B_y=v_y = 0$ and show transverse driving ($v_x$) at the lower (left hand panels) and upper (right hand panels) boundaries. For the current setup, at both boundaries we have $B_z > 0$ and in Fig. \ref{poynt_cartoon}, we show the different possibilities for the case $B_x < 0$. Ultimately, at either boundary, when the driving is increasing the magnitude of the transverse component of the field ($\left|B_x\right|$), it is injecting energy into the system. Otherwise, if the magnitude of the transverse component of the field is decreased by the driving, the imposed velocity will remove energy from the system. This argument extends naturally to the case with $v_y \ne 0$.

As each of the velocity components in the driver definition (eq. \ref{define_driver}) are independent from each other, when any given term is switched on, it is as likely to remove energy from the system as it is to inject new energy. It is only the continual action of each driver component that allows a net influx of energy. For example, a driver component acting in the positive $x$ direction on the lower $z$ boundary will induce a negative $B_x$ term here (see upper left hand panel of Fig. \ref{poynt_cartoon}) and thus begin to inject energy into the system. Therefore, in general, the longer each driving component acts for, the greater the Poynting flux it will drive into the simulation domain. As such, we can expect the long time scale cases (DC driving) to have a larger time-averaged Poynting flux than the short time scale cases (AC driving). This phenomenon is fundamentally responsible for the results that follow.

\begin{figure}[h]
  \centering
  \includegraphics[width=0.49\textwidth]{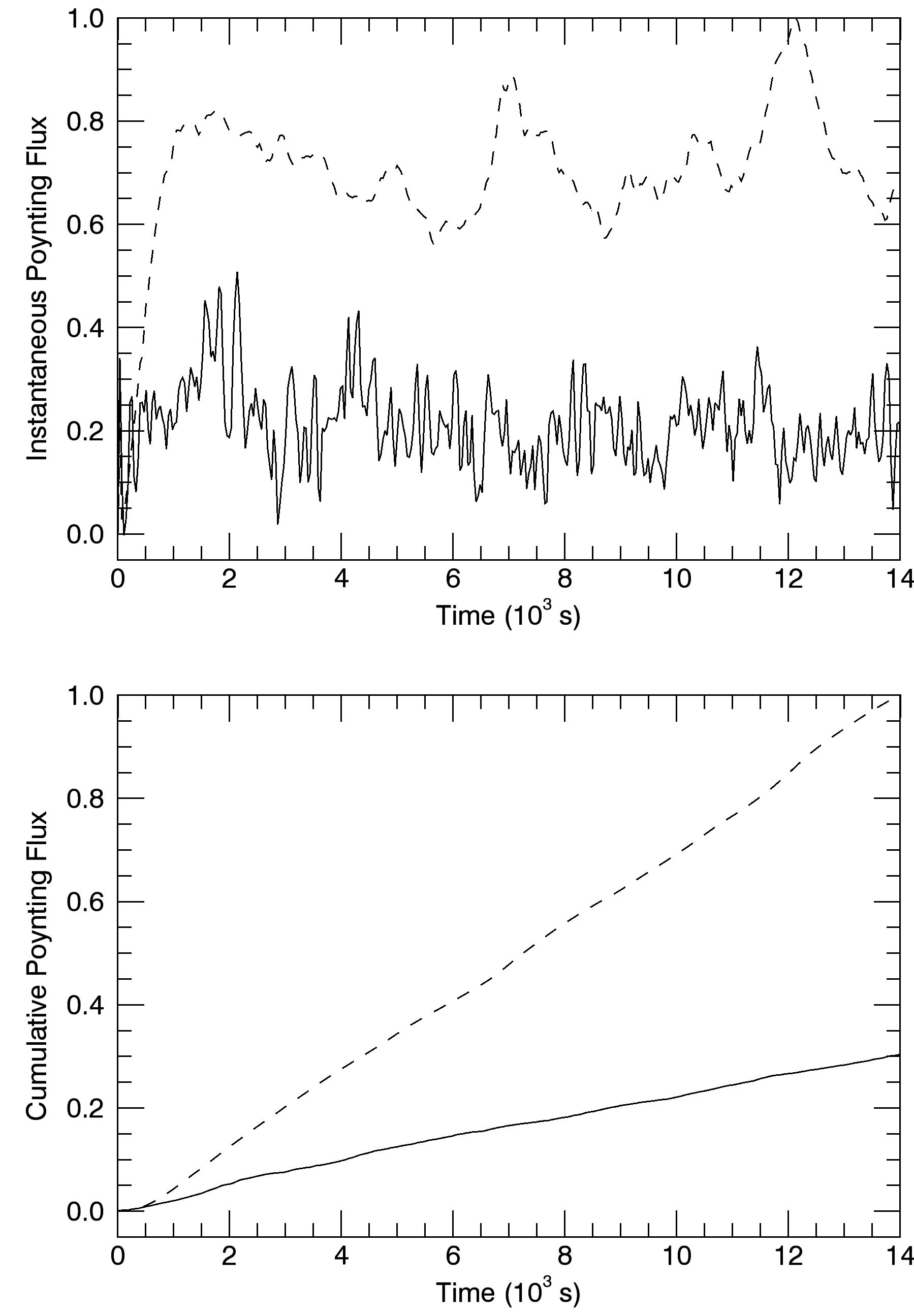}
  \caption{Poynting flux injected by medium amplitude AC (solid) and DC (dashed) driving. In the upper panel we show the instantaneous flux and in the lower panel we show the cumulative energy injection. In both cases, we have normalised the variables by the maximum of the dashed lines.}
  \label{poynt_plots}
\end{figure}

In Fig. \ref{poynt_plots}, we show the evolution of the Poynting flux for medium amplitude AC (solid) and DC (dashed) driving. In the upper panel, we show the instantaneous Poynting flux (equation \ref{Poynt_flux_exp}) and in the lower panel we show the cumulative Poynting flux. This second plot is the total energy injected since the start of each simulation and is simply the time integral of the lines in the upper panel. For both simulations, we see that after an initial increase, the instantaneous Poynting flux reaches a relatively steady state (when viewed over long enough time periods). As the mean value is positive in both cases, this ensures that there is a constant flux of energy injected into each of the simulations. However, over the experiment run-time, this does not necessarily lead to increasingly energetic systems as the radiative loss term is effective at removing energy, particularly in the dense chromosphere.

For the reasons discussed above, we see considerably higher energy injection rates for the DC driving (dashed lines) than for the AC driving (solid lines). We also see that the Poynting flux fluctuates rapidly in the AC simulations and much more slowly in the DC cases. This is a direct effect of the temporal variability associated with the different driving profiles. Despite this, the cumulative Poynting flux (lower panel) increases approximately linearly for both simulations.

\begin{figure}[h]
  \centering
  \includegraphics[width=0.49\textwidth]{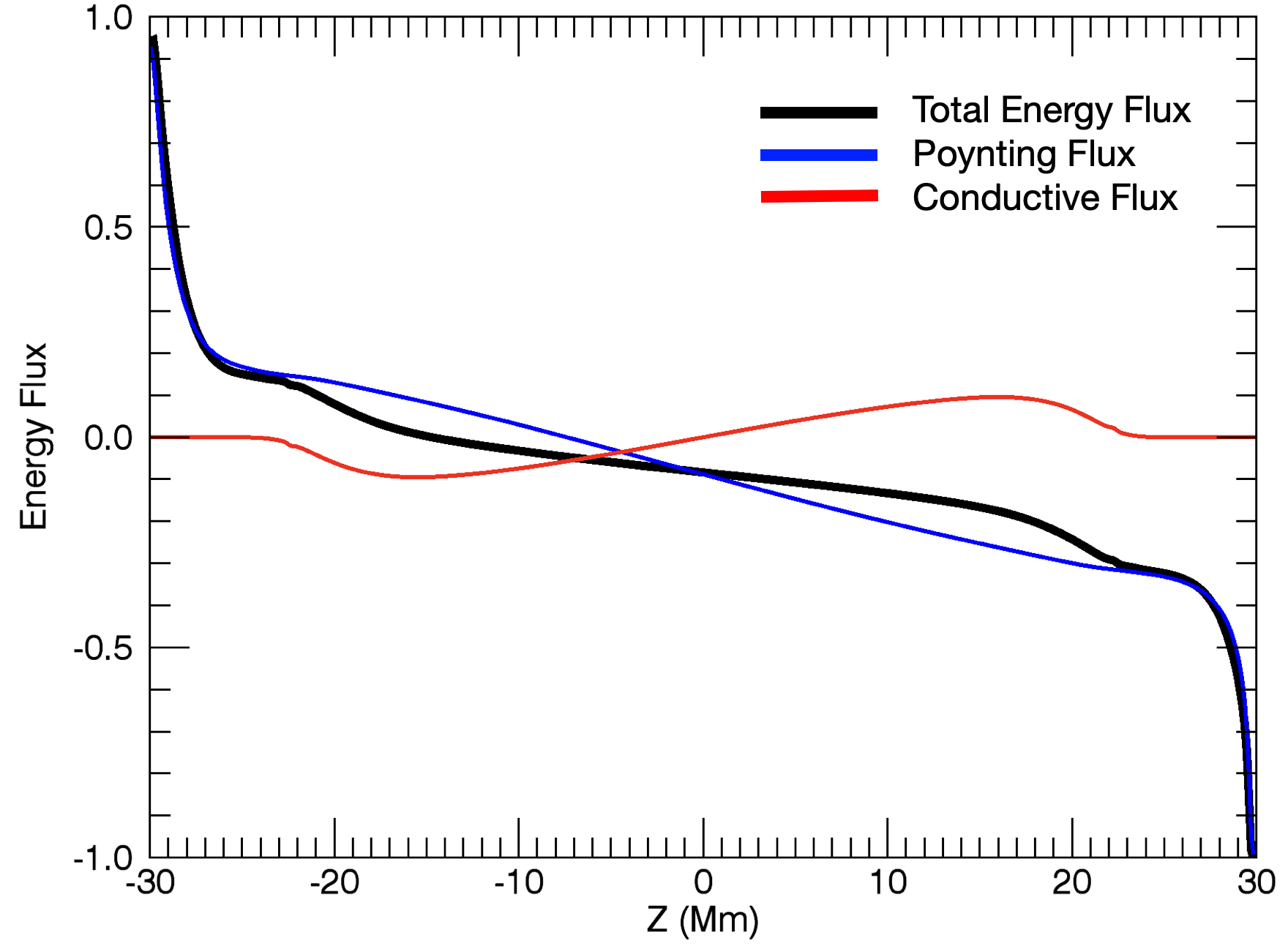}
  \caption{Vertical energy flux within the simulation domain. For the medium amplitude AC simulation, we show the temporally- and spatially- (in $x$ and $y$) averaged total flux (black), Poynting flux (blue) and conductive flux (red).}
  \label{flux_Z}
\end{figure}

Once energy is injected into the simulation domains, it is transmitted through the volume via a series of flux terms; namely, Poynting flux, kinetic energy flux, enthalpy flux, gravitational potential flux and conductive flux. In Fig, \ref{flux_Z}, we show the combined effects of these terms on the transfer of energy in the $z$ direction. To create this plot, we have averaged the $z$ component of each energy flux term in the $x$ and $y$ directions and over the duration of the simulation. We have then summed each term to give the mean combined flux parallel to the $z$ axis (black line). Here, a positive value means energy is flowing in the positive $z$ direction. We also show the contribution from the Poynting flux (blue line) and the conductive flux (red line). For all three curves, we have normalised by the maximum of the absolute value of the total energy flux (black line).

We see that throughout much of the domain, the energy transfer is dominated by the Poynting flux (compare black and blue lines). However, in the corona and transition region, the conductive flux also acts as an important conduit for transporting energy within the simulation volume. This relates to a coronal cooling term which transfers energy from hot plasma at the loop apex to the cooler layers of the atmosphere below. As the chromosphere is isothermal, there is no conductive flux within this layer. 

It is important to note that the rapid decrease in the absolute value of the energy flux for $|z| \gtrsim$ 26 Mm means that much of the energy is not transmitted to the corona in the central portion of the domain. Indeed only about 20 \% of the injected energy passes into the upper chromosphere. Some of the remaining energy is stored in the perturbed magnetic field, however, much is lost through radiative effects. For example, this can include slow wave chromospheric damping. In reality, whilst a large proportion of mechanical energy will not be passed into the corona, we cannot infer an accurate estimate for the transmission rate from these simulations due to the lack of chromospheric physics included in our model.

\begin{figure*}[h]
  \centering
  \includegraphics[width=0.97\textwidth]{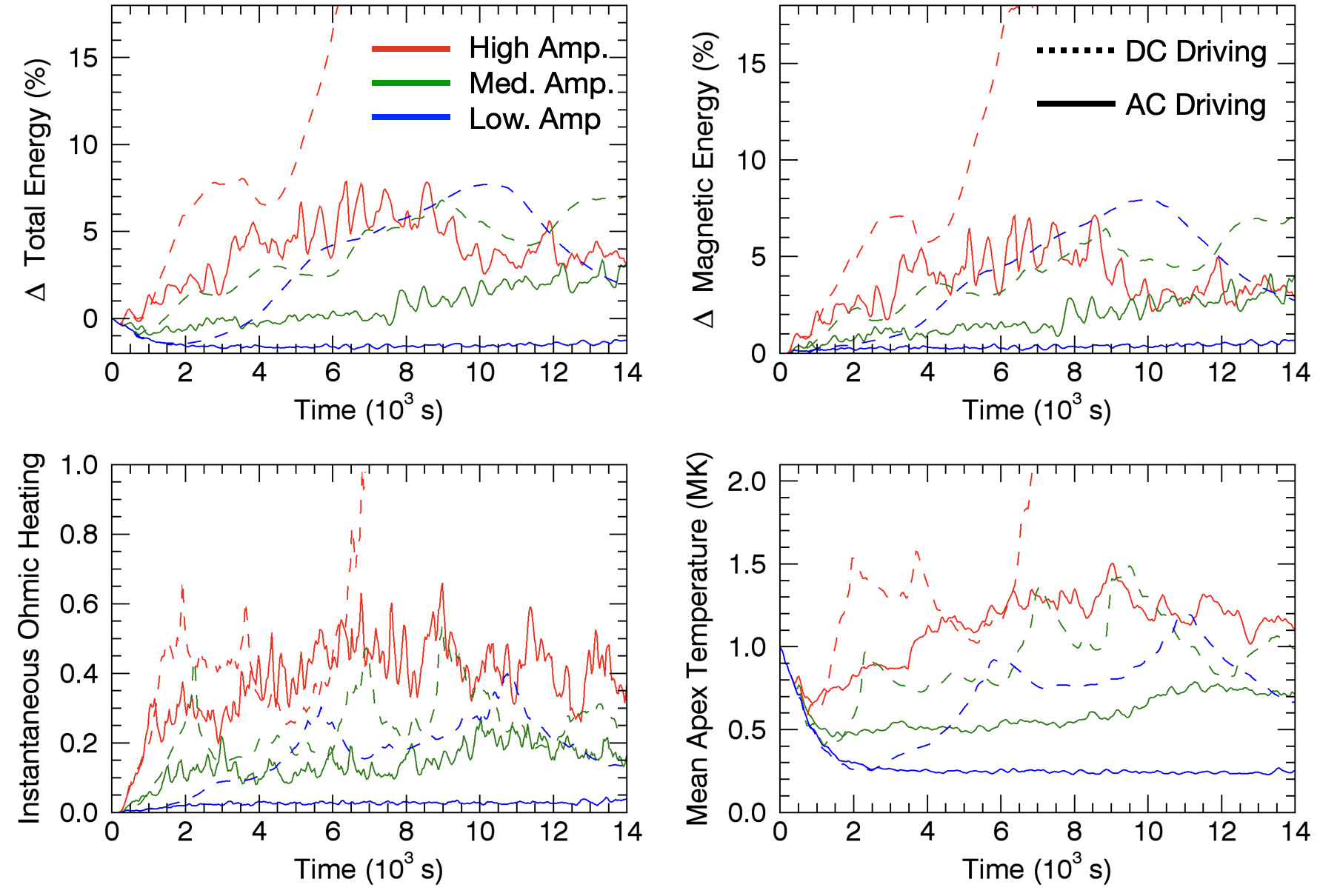}
  \caption{Volume integrated quantities for all six simulations. In the upper two panels, we show the change in the volume integrated total energy (left) and magnetic energy (right). In both cases, we show the percentage relative to the initial total energy. We also show the volume integrated Ohmic heating (lower left), normalised to the maximum rate in the high amplitude, DC simulation. Finally, we show the maximum temperature at $z=0$ in the lower right panel. In all figures, solid lines refer to AC cases and dashed lines refer to DC cases. Each colour refers to a different driver amplitude. }
  \label{vol_energetics}
\end{figure*}

\subsection{Energetics and thermal evolution}
The increased Poynting flux in DC driving cases provides a larger source of energy that can be dissipated in order to heat the plasma. In this section, we compare the effects of driving time scales and amplitudes on the components of the volume-integrated energy and on the temperature of the coronal plasma.

In the upper left panel of Fig, \ref{vol_energetics}, we show the change in the volume integrated energy for all six of the simulations. The change is shown as a percentage of the initial value, which is the same in all cases. For all panels in this figure, solid lines correspond to AC simulations and dashed lines correspond to DC cases. The different colours show the different driving amplitudes. The red, green and blue lines show high, medium and low amplitude cases, respectively. We note that the high amplitude DC simulation was run for a shorter time (approximately 7000 s) than the other simulations, as the computational time step became prohibitively short beyond this point. 

As expected, we typically see that higher amplitude driving is associated with a larger increase in the total energy. However, the random nature of the imposed velocity means that this is not necessarily true at all times (e.g. compare dashed blue and green lines). In most cases, the total energy increases during the course of the simulation. Despite this, the low amplitude AC driver (solid blue line) does not inject enough energy to balance the thermal losses (radiation) from the simulation. All simulations show an initial decrease in the total energy, although this occurs over a short time period ($< 500$ s) in the most energetic cases (red lines). This is because thermal losses are important from the simulation onset, whereas it takes some time for the drivers to begin to inject significant Poynting flux. In agreement with the discussion in the previous section, we see that, for all amplitudes, the DC driving consistently injects more energy than the AC driving (compare solid and dashed lines). Further, the characteristic time scales of each driver can be identified in this plot. In particular, the AC curves (solid lines) exhibit higher frequency variation than the DC cases.

In the upper right panel of Fig, \ref{vol_energetics}, we restrict our attention to the magnetic component of the energy. Again, the lines are normalised by the maximum of the initial (total) energy. By comparing the curves with the previous panel, we see that the majority of the energy increase is stored in the magnetic field. The initial, uniform field represents the minimum energy state for the amount of magnetic flux that passes through the simulation domain. As no flux is removed (or indeed added) to the simulation, the magnetic energy can only increase from the initial state. As such, the change in magnetic energy (relative to the initial value) is never negative in any of the simulations. The injected magnetic energy is associated with a non-potential component of the field which can be dissipated by resistive effects. 

The lower left hand panel of Fig. \ref{vol_energetics} shows the Ohmic heating produced by the dissipation of currents in the field. Here, we have normalised all curves by the maximum of the dashed red curve. As a direct result of the increased energy injection, we see more heating in higher amplitude cases and in the DC simulations. Despite integrating over the entire simulation volume, the Ohmic heating rate shows high temporal variation, particularly in the AC cases. This suggests that the energy release is inherently intermittent and bursty. However, we note that in all cases, there is a background threshold that the heating never falls below. As such, there is a steady component to this heating, at least when computed over a sufficiently large volume. It is likely that this background threshold is a function of the transport coefficients and may be much lower in higher Reynolds number plasmas.

In the lower right hand panel of Fig. \ref{vol_energetics}, we show the mean temperatures in the $z=0$ Mm plane (effectively the top of the corona) during all simulations. Unsurprisingly, we see that the simulations with higher Ohmic heating rates produce hotter coronae. In particular, DC driving cases support hotter atmospheres than their AC driving counterparts. It is also of note that the short time variations observed in the AC Ohmic heating profiles (solid lines in the lower left panel) are reduced in the thermal evolution. As with the comparison between the current and temperature in Fig. \ref{horiz_contour}, this is largely because the plasma does not cool instantaneously following a heating event. Consequently, the temporal gradients are reduced.

\begin{figure}[h]
  \centering
  \includegraphics[width=0.49\textwidth]{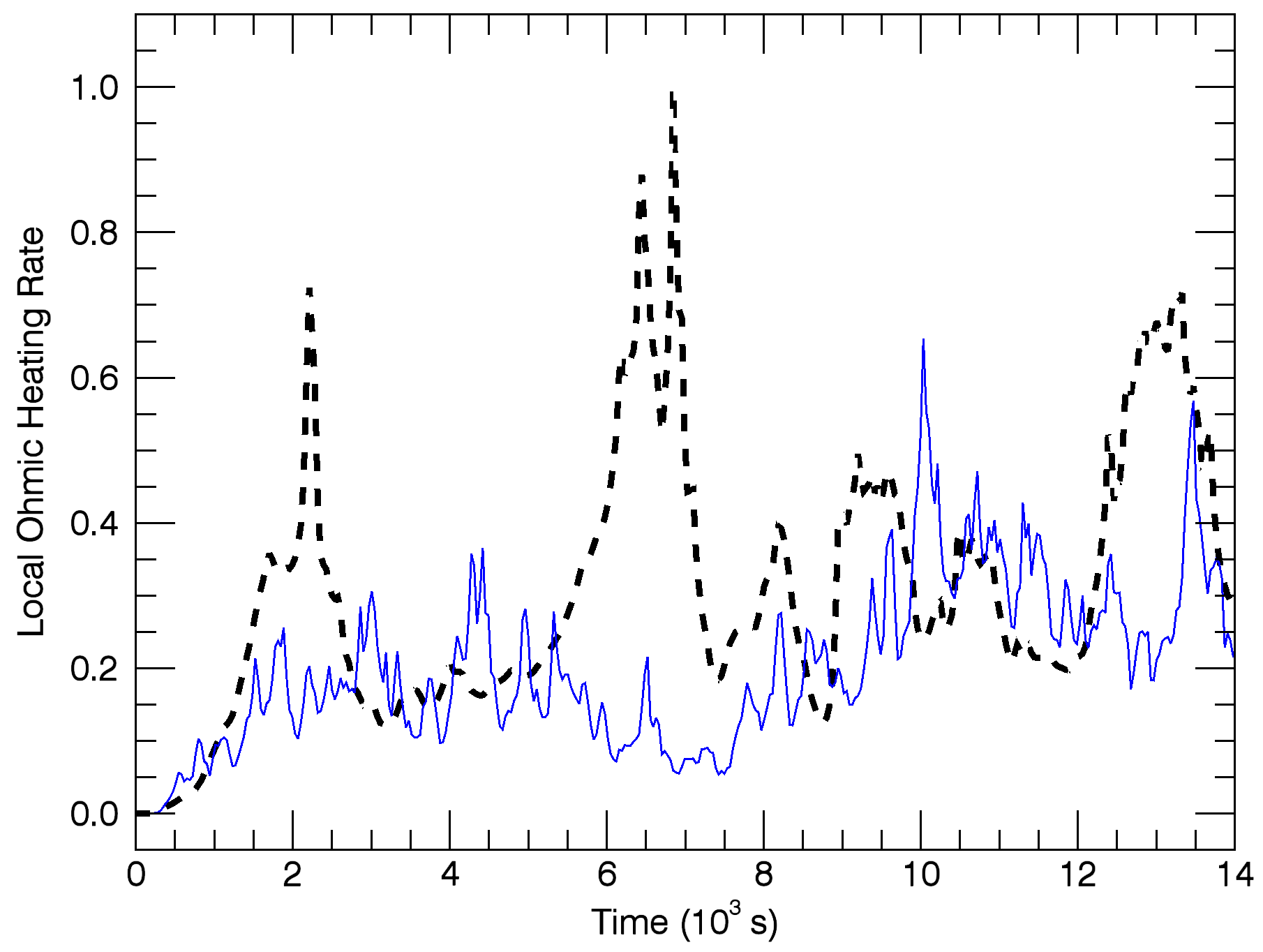}
  \caption{The volume integrated Ohmic heating rate in $-1 \le x, y, z \le 1$ Mm for the medium amplitude AC (solid blue) and DC (dashed black) simulations. Here we have normalised both lines by the maximum of the DC case (black). }
  \label{local_ohm_heat}
\end{figure}

As much of the local variation is lost by averaging over the entire simulation volume, in Fig. \ref{local_ohm_heat}, we show the local Ohmic heating rate as a function of time in a small sub-volume given by $-1 \le x, y, z \le 1$ Mm. For clarity, we only show the medium amplitude AC (solid blue line) and DC (dashed black line cases). In both cases, we have normalised the curves by the maximum of the dashed black line. As expected, the mean of the DC heating rate is larger, however, it only exceeds the AC heating rate at a few discrete, large scale events. In both cases, the energy release is inherently bursty, although again, there is a minimum threshold which effectively acts as a steady heating term. The DC heating events tend to be larger and more long-lived but also less frequent than the AC events. The importance of the frequency of heating is reviewed in \citet{Reale2010} and means we should expect that more of the cooling phase will be observed for DC heating. 

\subsection{Field line analysis}
As the thermodynamic evolution of coronal plasma is typically dominated by field-aligned conduction, the response of the solar atmosphere to heating events is frequently studied using 1D modelling. The complexity of these simulations prohibits any single field line being tracked throughout the experiments. However, we are able to provide insight into the characteristic behaviour of different field lines by sampling many field lines at different times. In the following, we have tracked $10^4$ field lines with foot points uniformly distributed on the lower boundary at all output times during the simulation. The starting points do not move in time and thus, as the field is advected by the imposed drivers, we have not tracked the same field lines. The simulations produce 400 data outputs and thus we have a sample of $4 \times 10^6$ field lines from each experiment. 

\begin{figure}[h]
  \centering
  \includegraphics[width=0.49\textwidth]{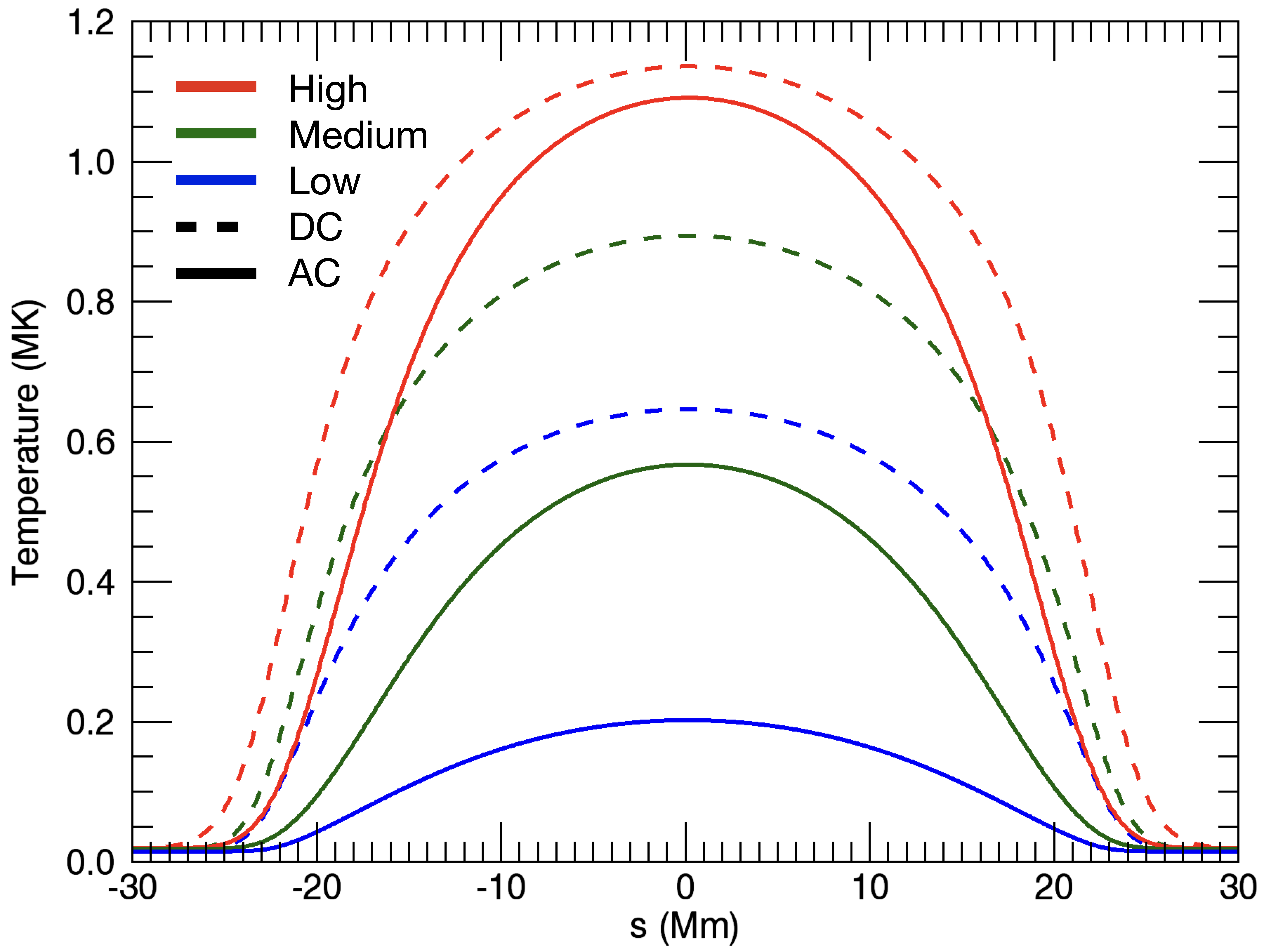}
  \caption{Mean temperature profile along field lines from all simulations. Solid lines show AC cases and dashed lines show DC cases. Each colour refers to different driver amplitudes. For each simulation, the averaging was calculated for $10^4$ different field lines sampled every 35 s.}
  \label{vert_temp_str}
\end{figure}

In Fig. \ref{vert_temp_str}, we show the average temperature profile along these field lines for the different simulations. As the field lines are different lengths, we have only considered the central 60 Mm of each field line (the minimum possible length). The variable, $s$, measures the distance along each field line and $s = 0$, is defined to be at $z=0$ Mm. All six simulations are shown with dashed and solid lines referring to AC and DC driving, respectively and red, green and blue corresponding to high, medium and low amplitude driving. The simulations with the higher apex temperatures (lower right hand panel of Fig. \ref{vol_energetics}) are hotter throughout the coronal volume but all cases retain the cool chromospheres at the temperature floor of $2 \times 10^4$ K. This is because the high density here means that radiative losses are always in excess of any heat that conducts from the corona. However, it is apparent that the hotter simulations support an extended coronal volume as the transition regions move closer to the simulation boundaries. 

\begin{figure}[h]
  \centering
  \includegraphics[width=0.49\textwidth]{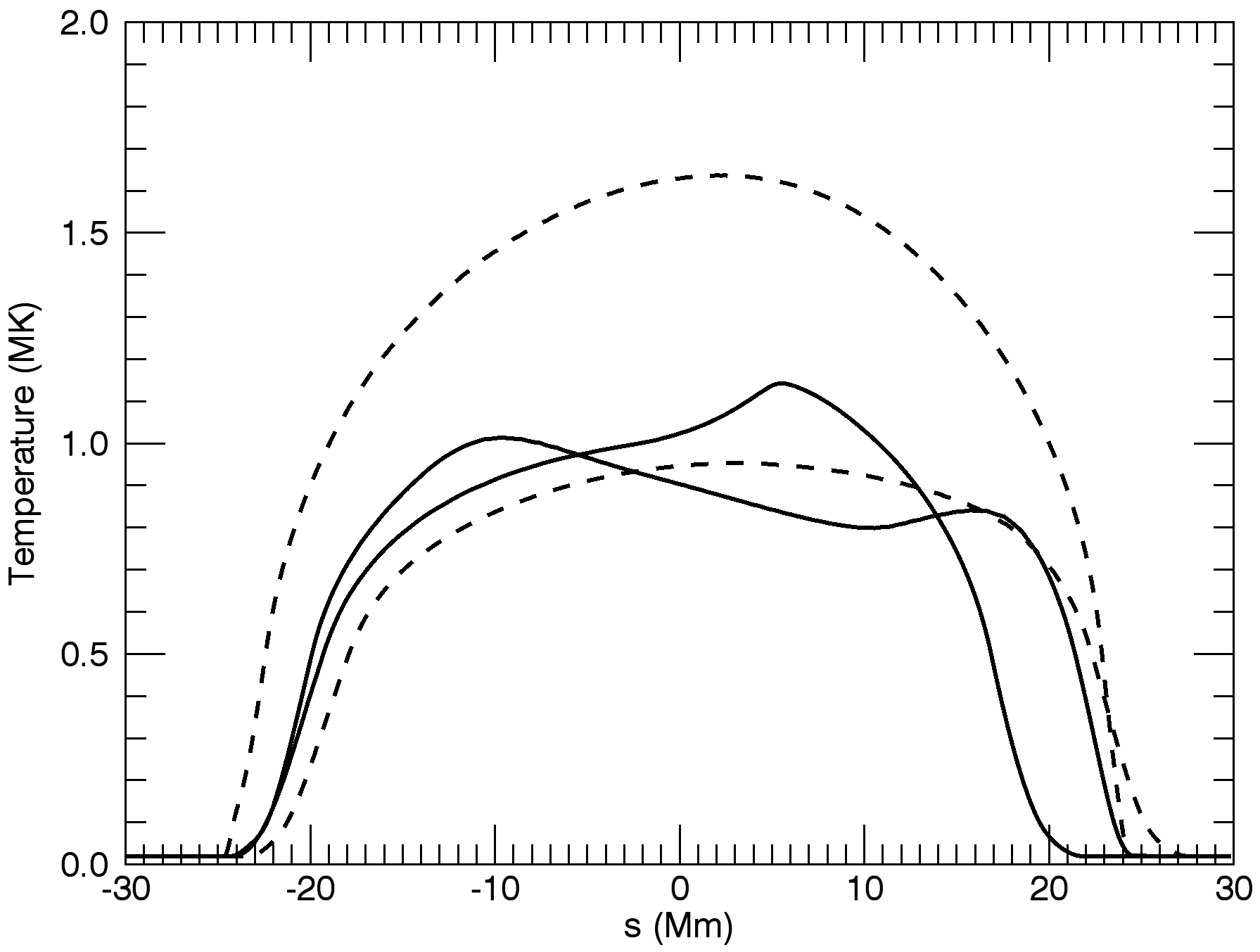}
  \caption{Temperature profile along two field lines from all the medium amplitude AC (solid lines) and DC (dashed lines) simulations.}
  \label{vert_temp_ind}
\end{figure}

We note that the large temperature gradients that might be expected in the transition regions are not visible in Fig. \ref{vert_temp_str}. This is partially due to the numerical treatment of conduction employed here (see Sect. \ref{num_method}), but is also caused by averaging over many field lines. Each field line has transition regions at different $z$ values, according to its own thermal evolution. As such, the wide spread of transition region locations produces the relatively shallow average temperature profiles. In Fig. \ref{vert_temp_ind}, we show the temperature profile of field lines in the medium amplitude AC (solid lines) and DC (dashed line) simulations. Here, the temperature jumps in the transition regions are somewhat steeper (compare to green lines in Fig. \ref{vert_temp_str}) although they are still reduced by the adjustment to thermal conduction. Although the DC simulations are in general hotter, as we see here, this is not necessarily the case for all field lines. As mentioned earlier, the DC heating events tend to have lower frequencies and thus allow more time for field lines to cool than in the AC simulations. We also note that the AC temperature profiles (solid lines) are more asymmetric than for the DC cases. This is a consistent trend across field lines in the simulations and is discussed in more detail below.

\begin{figure*}[h]
  \centering
  \includegraphics[width=0.97\textwidth]{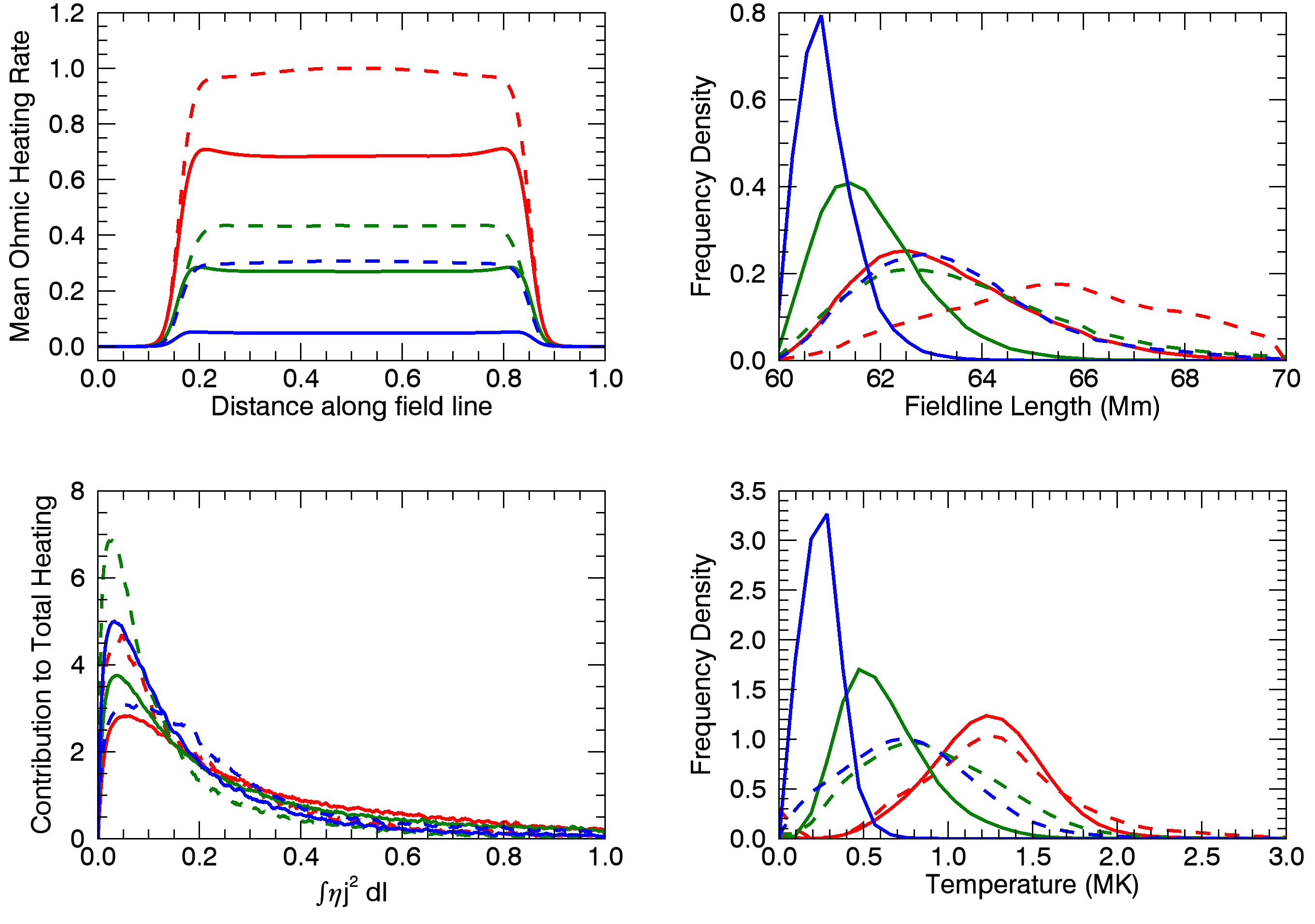}
  \caption{Properties of field lines sampled from all six simulations. \emph{Upper left:} The mean Ohmic heating rate along field lines. \emph{Upper right:} The distribution of field line lengths.  \emph{Lower left:} The contribution of field lines with different Ohmic heating rates to the total energy release. \emph{Lower right:} The distribution of field line temperatures. In the upper left panel, curves are normalised to the maximum of the dashed red line. The lines in all other panels are normalised such the area under each curve is unity.} 
  \label{fieldline_plots}
\end{figure*}

In Fig. \ref{fieldline_plots}, we show some general properties of the field lines traced in the simulation results. In the upper left hand panel, we show the mean Ohmic heating profile along all of the sampled field lines in each simulation. To allow for different length field lines, we have normalised all lengths to unity ($x$ axis) before computing the average. We have also normalised the value of the mean Ohmic heating rates to the maximum of the red dashed curve. In all cases the heating rate falls to zero close to the upper and lower boundaries because the resistivity is set to zero here (eq. \ref{eq:eta}). Unsurprisingly, we see that the more energetic simulations produce higher Ohmic heating along the entire coronal section of the field lines. In all cases, the average heating rate is approximately constant across this coronal volume. This is largely a result of the initial uniform magnetic field and is likely to change if the field strength changes in height \citep[for example in an arcade structure as used in][]{Howson2020}. Additionally, despite this average uniformity, we note that heating on individual field lines can be highly localised and thus non-uniform. 

In the upper right hand panel of Fig. \ref{fieldline_plots}, we show the distribution of field line lengths from each simulation. As the height of the box is 60 Mm and field lines connect to opposite boundaries, the minimum length of a field line in all cases is 60 Mm. Indeed, all field lines initially have this length. However, as the driving progresses, and the magnetic field becomes stressed, the length of field lines increases as different field lines take increasingly convoluted paths between the upper and lower boundaries. Both higher amplitude and longer time scale driving increase the mean length of field lines more than lower amplitude and/or shorter time scale driving. This is because individual components of the imposed velocities (see eqs. \ref{vx_def} \& \ref{vy_def}) will displace magnetic foot points by a greater distance if the amplitude is larger or they are active for longer (increased time scale). We note that this also affects the width of the field line length distributions, with more energetic simulations producing a greater spread. The greater field line lengths in DC and high amplitude simulations will promote apex temperature increases due to a decrease in thermal conductive losses to the lower atmosphere. However, as the mean length increase is relatively small, this effect will be negligible in comparison to the increased energy injection for these simulations.

\begin{figure*}[h]
  \centering
  \includegraphics[width=\textwidth]{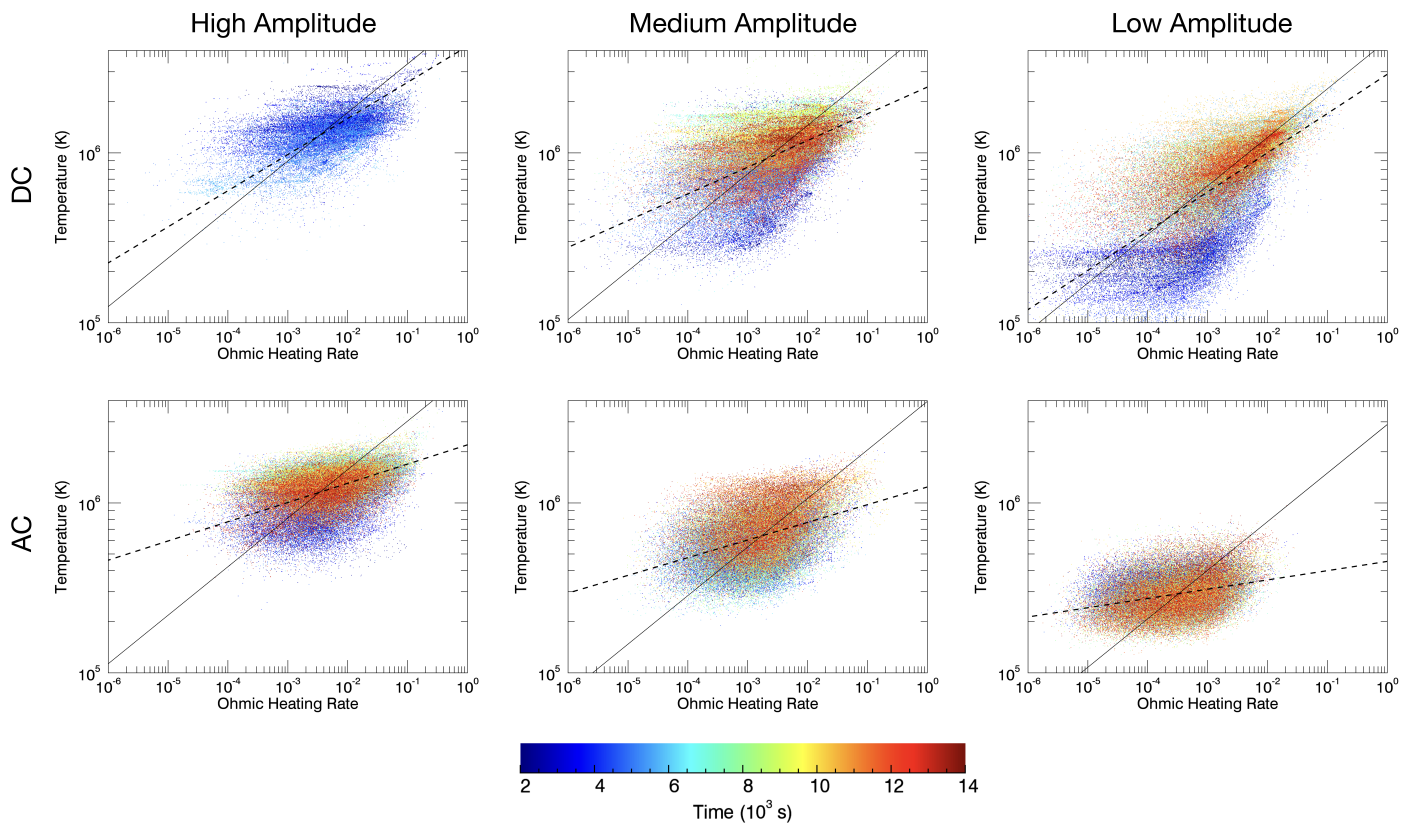}
  \caption{Maximum temperature against the integrated Ohmic heating rate along the length of each field line. The heating rate is normalised to the maximum observed in the high amplitude DC simulation. Each panel shows the results of a different simulation. Individual points each show different field lines and the colour shows the simulation time when the field line was traced. Solid lines show the expected scaling law assuming thermodynamic equilibrium. Dashed lines show the scaling measured in each simulation.} 
  \label{temp_vs_ohm}
\end{figure*}

In the lower left hand panel of Fig. \ref{fieldline_plots}, we show that the total heating is mainly composed of heating from many field lines with weak heating rather than the few field lines with strong heating. To generate this plot, we grouped field lines according to the magnitude of their heating ($x$ axis normalised by the maximum heating rate in each simulation). Then, we calculated the total heating produced by field lines in each group ($y$ axis). Whilst the exact shape of the distribution differs between simulations, it is clear that in all cases, energy release is dominated by relatively weak, but numerous, events.

In the lower right hanx panel of Fig. \ref{fieldline_plots}, we show the distribution of the maximum temperature of the sampled field lines. Once again, we see that the more energetic simulations produce more field lines with higher temperatures. Additionally, we note that the temperature distributions are positively skewed with a large number of cool field lines and a relatively small number of hotter field lines. This is associated with the Ohmic heating distributions (lower left panel), which show that relatively few field lines have high heating rates and thus not many field lines reach the very high temperatures.

Although the majority of energy release is associated with low energy events, the highest temperatures in each simulation are only attained where the Ohmic heating rate is very high. In Fig. \ref{temp_vs_ohm}, we show how the maximum temperature on each of the traced field lines compares to the total Ohmic heating on that particular field line (each point represents an individual field line). For clarity, we only show a subsample of the field lines by selecting every tenth snapshot from each simulation. We also omit the first 2000 seconds of evolution as many of the field lines have very similar characteristics at early times due to the uniformity of the initial conditions. In each panel, we show the result of a different simulation and the colours of the points refer to the simulation time at which the field line was traced. The solid line in each panel shows the scaling predicted by the RTV scaling law \citep{Rosner1978}, under the assumption that all field lines have the same length. The dashed line shows the actual scaling found in each simulation. Whilst the assumption on field line length is violated, the fractional variation is small and this is not a large factor in the disagreement between the solid and dashed lines. Instead, the discrepancy is a result of field lines not being in thermodynamic equilibrium when the measurement is taken. In general, the DC simulations (upper row) show better agreement with the predicted scaling law, because their evolution is much less dynamic and thus remains closer to the equilibrium state. In all cases, the observed power law is less steep than expected (compare gradients of dashed and solid lines). This is largely because previously heated field lines (where the Ohmic heating is now reduced) can still show enhanced temperatures, as they do not cool instantaneously.

In agreement with previous plots, we see that the DC and higher amplitude simulations exhibit hotter temperatures than other cases. We also see that the DC simulations (except the high amplitude case with the reduced run time) show a wider distribution in terms of the field line apex temperatures (also see lower right hand panel of Fig. \ref{fieldline_plots}) than their AC counterparts. Further, as the colour of each point refers to the simulation time, the temporal evolution can be tracked by moving from blue to green to red points. The upper left hand panel (high amplitude DC simulation) only contains blue points as the simulation run time is much shorter than for other cases. In all other panels, the difference between early and late times can be seen by comparing blue and red points. We see that blue points (early times) generally lie at lower temperatures and lower Ohmic heating rates as it takes time for the drivers to produce the small scales required for efficient energy release. As such, the simulations are cooling at early times (see also lower right hand panel of Fig. \ref{vol_energetics}). In all cases, the energy release requires complexity to be injected by the imposed velocity drivers. Therefore, it typically requires many different Gaussian components to have been activated (eqs. \ref{vx_def} \& \ref{vy_def}) before the highest heating rates develop. This takes longer for long time scale drivers and, as such, we see a more distinct population of blue points in the DC panels (medium and low amplitudes) than in the AC cases. The only exception to this behaviour is in the low amplitude AC case (lower right panel). This simulation never reaches a heating phase as the Poynting flux is insufficient to overcome the thermal losses.

\begin{figure}[h]
  \centering
  \includegraphics[width=0.49\textwidth]{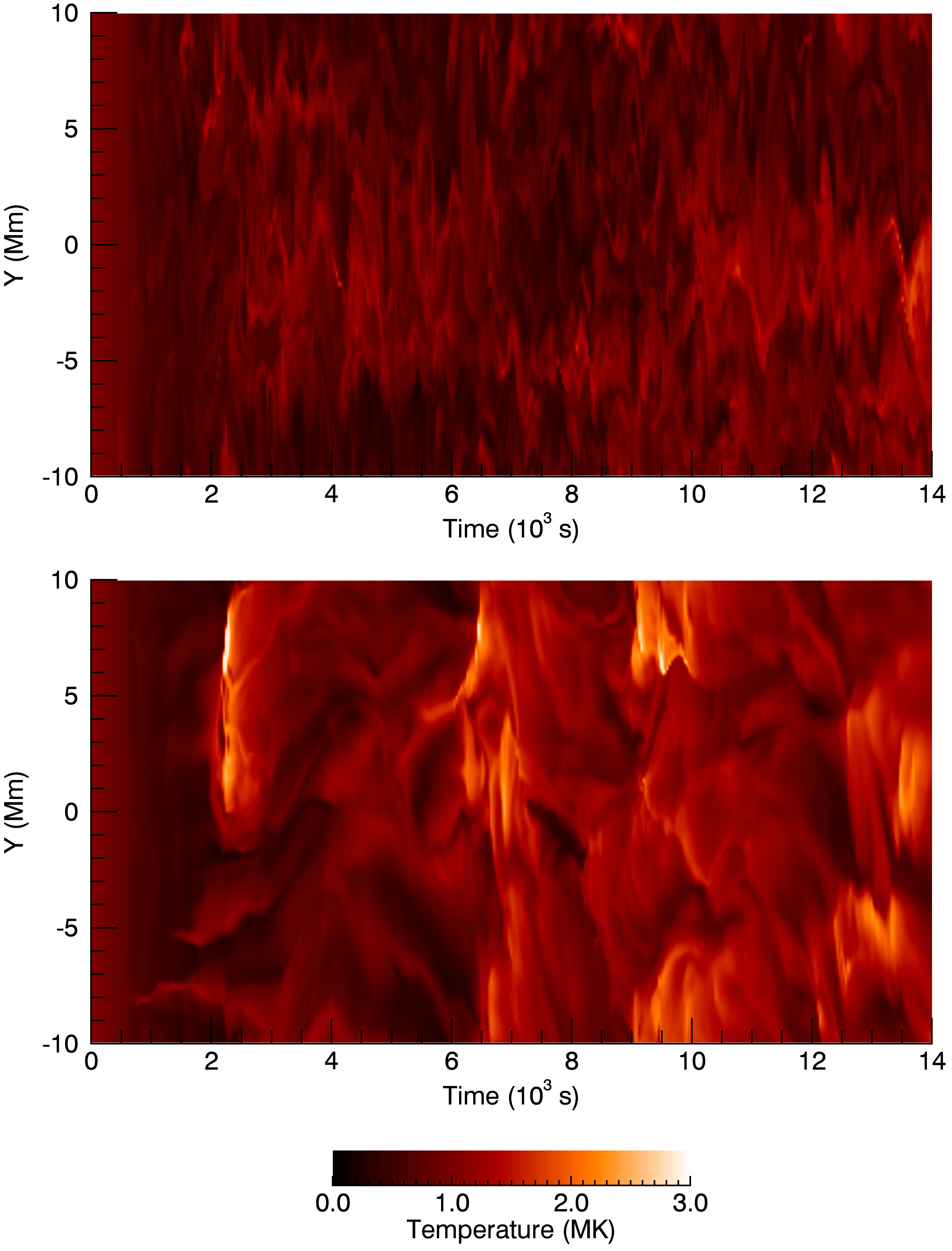}
  \caption{Temporal evolution of the temperature in the line $x=z=0$ Mm for the medium amplitude AC (upper) and DC (lower) simulations.}
  \label{temp_time}
\end{figure}

\subsection{Distinguishing features of AC and DC heating}

The different characteristic time scales of the velocity driving translate into increased dynamism for the AC simulations within the coronal fraction of the atmosphere. This can be seen in short time scale variation in the velocity field and the density, for example. As such, synthetic observables generated by these simulations can be used to easily distinguish between the two driving time scales \citep[][]{Fyfe2021}. This enhanced dynamism can also be observed in the evolution of the apex temperature which is affected by the different heating characteristics and the different advection profiles for the AC and DC cases. In Fig. \ref{temp_time}, we show the evolution of the temperature in the line $x=z=0$ Mm for the medium amplitude AC (upper) and DC (lower) simulations. In the upper panel, we see much higher frequency variation, lower mean temperatures and no large scale, high temperature formation as can be seen in the lower panel. This highlights the fundamental differences between the AC and DC evolution. The infrequent but large heating events that characterise the energy release in the DC simulations are easily identifiable in the lower panel (e.g. for $y > 0$ Mm at $t \approx 2500$ s). Whilst a wide range of frequencies coexist in the Sun's atmosphere, the differences between the temperature evolution, suggests that these heating mechanisms may be distinguishable.

\begin{figure}[h]
  \centering
  \includegraphics[width=0.49\textwidth]{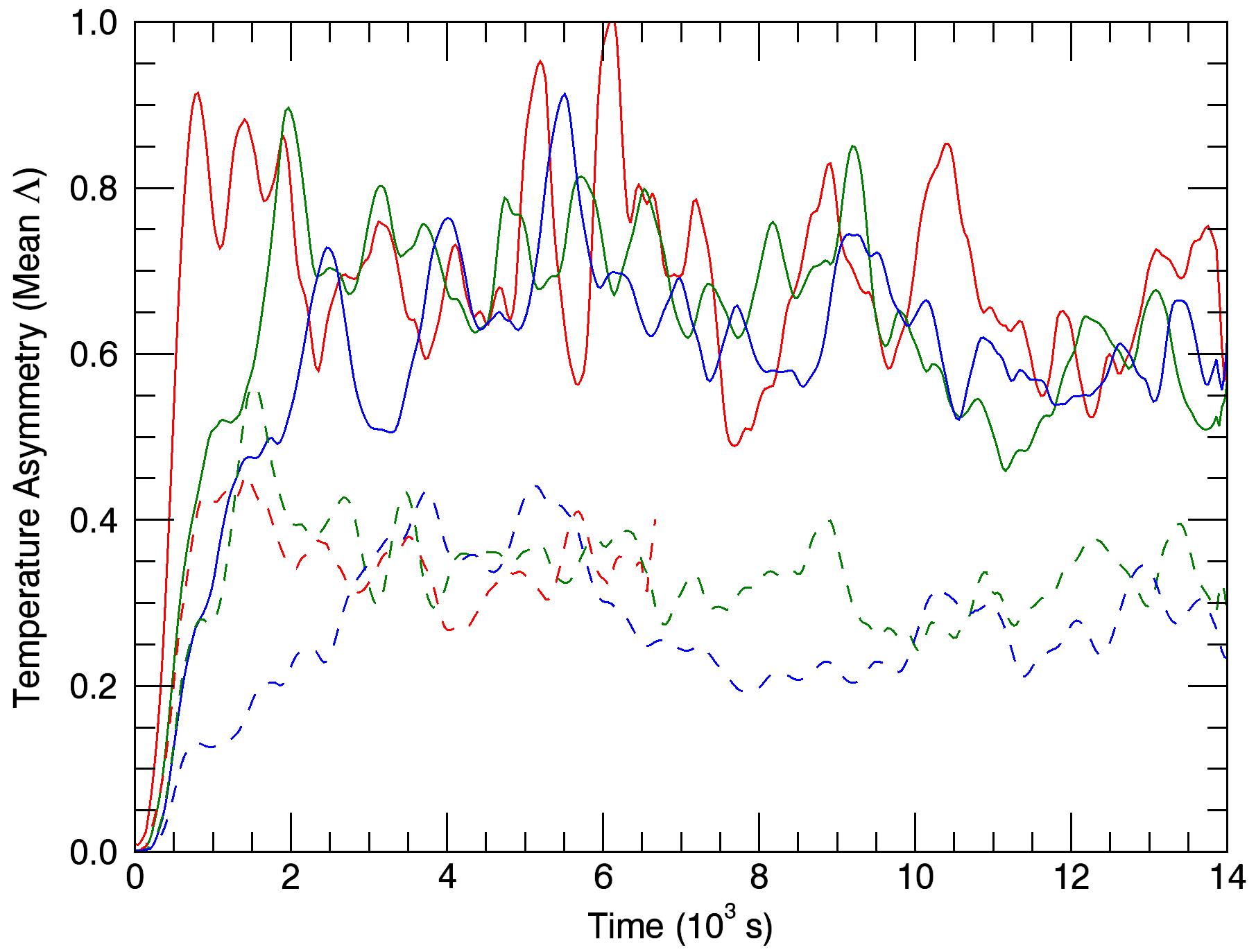}
  \caption{Evolution of the mean asymmetry, $\Lambda$, of temperature profiles along the traced field lines. Results are shown for high (red), medium (green) and low (blue) amplitude AC (solid) and DC (dashed) simulations.}
  \label{temp_asym}
\end{figure}

An additional characteristic which distinguishes between the heating in different simulations is identified by considering the asymmetry of temperature profiles along individual field lines. As mentioned previously, Fig, \ref{vert_temp_ind} shows two field lines from AC simulations (solid lines) which have more asymmetric temperature profiles than the two field lines from DC simulations (dashed lines). In order to show that this is typical for field lines in the different simulations, we define a measure of the temperature asymmetry on a given field line by
\begin{equation}
\Lambda = \frac{\left|\int T(s)  - \int T(-s) \right| \,\, \mathrm{d} s}{\int T(s) \, \, \mathrm{d} s} .
\label{eq_asym}
\end{equation} 
Here, $s$ paramaterises the field line. In this equation, the numerator represents the total difference between the temperature profile on either side of the loop apex. Hence, it is zero if the temperature profile is symmetric. The denominator then normalises the measure such that the asymmetry of field lines with different mean temperatures can be compared. 

In Fig. \ref{temp_asym}, we show the evolution of the mean (across all traced field lines) of $\Lambda$ for the high (red), medium (green) and low (blue) amplitude AC (solid) and DC (dashed) lines. We have normalised all curves by the maximum of the solid red line. We see that the AC simulations produce more asymmetric temperature profiles than the equivalent DC cases. After the initial adjustment phase of the simulations (the initial conditions are symmetric), the asymmetry remains roughly constant and the difference between the AC and DC cases is consistent. Additionally, we also note that this measure of the asymmetry in the field line temperature profiles is not affected by the amplitude of the velocity drivers in each simulation. Therefore, measuring the temperature asymmetry on coronal field lines may provide important information about the driving time scales that are most relevant for energy release.

\section{Discussion and Conclusions} \label{Discussion}
In this article, we have presented the results of a series of large scale MHD simulations of coronal heating in a gravitationally stratified solar atmosphere. Energy is injected by perpendicular velocity drivers imposed at magnetic foot points at both the upper and lower boundaries of the computational domains. The fundamental characteristics (e.g. amplitude, length scales, time scales) of these drivers are selected from statistical distributions to produce random motions. These drivers are designed to mimic the photospheric flows which inject energy into the Sun's atmosphere. 

From a parameter study on the driving time scales and on the velocity amplitudes, we find that longer driving time scales and higher velocities inject a greater Poynting flux into the system, produce larger currents and ultimately dissipate more energy as heat. Typically, the simulations evolve towards a steady state where the rate of energy dissipation in the corona is balanced by the loss mechanisms (e.g. radiation and conduction to lower layers of the atmosphere). Heating in DC simulations is characterised by larger scale and longer lasting energy release events that occur more infrequently than in AC experiments.

In contrast to many existing models of AC heating, the short time scale driving in this setup stores large quantities of magnetic energy in the coronal field. This typically enhances the Poynting flux and permits greater energy injection than is possible in simulations where the driver only perturbs the field about an initial equilibrium. As such, even in the AC simulations (medium or high amplitude), the rate of energy injection is sufficient to balance coronal losses. Despite this, as with all large scale 3D MHD simulations, the viscous and magnetic Reynolds numbers that can be obtained here are orders of magnitude smaller than might be expected in the solar corona. As a result, it is uncertain whether the energy injected by these drivers would be dissipated on sufficiently short time scales to be relevant for coronal heating. Further, it remains unclear whether the relative efficiency of AC and DC heating is constant across a large range of different Reynolds numbers.

Another limitation of the current model is that it does not include a full and accurate treatment of chromospheric physics. Indeed, in these simulations the lower layers of the atmosphere simply act as a mass reservoir from which plasma can evaporate in response to heating events. Despite this, and although the transmission of energy through the chromosphere remains poorly understood, coronal observations have measured oscillatory power at a range of different frequencies \citep[e.g.][]{Morton2015, Morton2019}. As such, we know that motions with similar characteristic time scales to the driving presented here are transmitted into the corona. In fact, these observational results detect significantly more power at lower frequencies (similar to DC driving) than at higher frequencies (AC driving). Whilst this may be partly due to observational biases; e.g. power at higher frequencies being difficult to observe due to smaller length scales and higher dissipation rates, it seems that there is significantly more power in low frequency modes. As our comparison, which assumes the same power for AC and DC modes, shows greater energy dissipation in the DC simulations, these observations suggest that this result may be enhanced further. As such, we expect DC heating mechanisms to dominate energy release in the closed corona.

The evolution induced by the random driving is inherently complex and thus, it is neither possible, nor particularly helpful, to analyse the nature of individual heating events. Instead, we have focussed on volume integrated and averaged quantities which provide information about the typical properties of plasma parameters (e.g. temperature). One such variable which shows a clear 
difference between AC and DC simulations is the asymmetry of the temperature profiles along magnetic field lines. In particular, AC driving typically produces temperature profiles which are, on average, more asymmetric than for DC driving. Furthermore, this difference is independent of the magnitude of the energy release. Therefore, whilst this parameter may be difficult to measure in observational studies, it may provide a useful insight into the fundamental nature of coronal heating.
\vspace{1cm}

{\emph{Acknowledgements.}} The research leading to these results has received funding from the UK Science and Technology Facilities Council (consolidated grant ST/N000609/1), the European Union Horizon 2020 research and innovation programme (grant agreement No. 647214). IDM received funding from the Research Council of Norway through its Centres of Excellence scheme, project number 262622. This work used the DiRAC Data Analytic system at the University of Cambridge, operated by the University of Cambridge High Performance Computing Service on behalf of the STFC DiRAC HPC Facility (www.dirac.ac.uk). This equipment was funded by BIS National E-infrastructure capital grant (ST/K001590/1), STFC capital grants ST/H008861/1 and ST/H00887X/1, and STFC DiRAC Operations grant ST/K00333X/1. DiRAC is part of the National e-Infrastructure.

\bibliographystyle{aa}        
\bibliography{Straight_Field.bib}           

\end{document}